\documentclass[prd,twocolumn,superscriptaddress,showpacs]{revtex4}
\usepackage{graphicx}
\usepackage{float}
\usepackage{dcolumn}
\usepackage{float}
\usepackage{bm}
\usepackage{mathrsfs}
\usepackage{ulem}
\usepackage[per-mode = fraction]{siunitx}
\usepackage{hyperref}

\usepackage{xcolor}
\usepackage{soul}
\usepackage{amssymb}
\usepackage{amsthm}
\usepackage{mathtools}

\newcommand{\be}{\begin{equation}}
\newcommand{\ee}{\end{equation}}
\newcommand{\ba}{\begin{eqnarray}}
\newcommand{\ea}{\end{eqnarray}}

\begin{document}

\title{Genesis and fading away of persistent currents in a Corbino disk geometry}
\author{Yuriy Yerin}
\affiliation{ Dipartimento di Fisica e Geologia, Universitá degli Studi di Perugia, Via Pascoli, 06123 Perugia, Italy}
\author{V.P.~Gusynin}
\affiliation{Bogolyubov Institute for Theoretical Physics, National Academy of Science of
Ukraine, 14-b Metrologicheskaya Street, Kiev, 03143, Ukraine}
\author{S.G.~Sharapov}
\affiliation{Bogolyubov Institute for Theoretical Physics, National Academy of Science of
Ukraine, 14-b Metrologicheskaya Street, Kiev, 03143, Ukraine}
\affiliation{Kyiv Academic University, 03142 Kyiv, Ukraine}
\author{A.A.~Varlamov}
\affiliation{CNR-SPIN, via del Fosso del Cavaliere 100, 00133, Rome, Italy}
\date{\today }

\begin{abstract}
The detailed analytical and numerical analysis of the electron spectrum, persistent currents, and their densities for an annulus placed in a constant magnetic field (Corbino disk geometry) is presented. 
We calculate the current density profiles and  study their dependence on the inner and outer radii  of the annular. We study evolution of the persistent currents and track their emergence and decay for different limiting cases of such a geometry, starting from a nanodot and ending by a macroscopic circle. Our analytical results for the currents are confirmed by the agreement between the integration of the corresponding current densities and the application  of the Byers-Yang formula, when it is applicable. 
Among other results we find the general expression for the persistent current in a narrow annulus, which in the one channel approximation reproduces the well-known result for quasi-one dimensional mesoscopic metallic ring. Moreover it allows to analyze the multi-channel case of a relatively wide annulus. Our study can be used for more accurate treatment and interpretation of the experimental data with measurements of the persistent currents in different doubly-connected systems. 
\end{abstract}

\maketitle

\section{Introduction}

While the Landau diamagnetism of free electron gas \cite{Landau} is often regarded as a standard
textbook knowledge, the discussion of the role of the edge states arising in finite systems is less known, although it is only one year younger \cite{Teller}. 
This analysis addressed the naturally risen question, why the Landau’s calculation of magnetization and similar treatment of the nondissipative transport coefficients remain correct for large enough system with finite boundaries  \cite{Heuser}. 
It is remarkably that these studies revealed the presence of the macroscopic nondissipative persistent edge currents flowing along the boundaries of the sample.

The persistent currents can also exist in doubly-connected systems due to the Aharonov-Bohm effect \cite{Aharonov1959PR}. 
For instance, it  was predicted \cite{Kulik1, Kulik2010} that in a hollow thin-walled normal metallic cylinder or ring with 
the small enough radius $R$ threaded by a magnetic flux $\Phi$ the persistent current can flow.  
Its magnitude oscillates as
$I  \sim (\left| e \right|{v_F}/R) \sin ({2\pi \Phi /{\Phi _0}})$, where $v_F$   
is the Fermi velocity and $\Phi_0  = h c/e$  is the magnetic flux quantum. 
The diamagnetic currents in the restricted geometry, including rings, were studied in between 60-70's e.g. in  \cite{Prange, Bogachek, Nedorezov}. It was also demonstrated that the account for such geometrical effects can lead to the magnetic response of the  magnitude  larger  than the Landau diamagnetic moment 
(see the reviews in Refs.~\cite{Richter, Gurevich}.)

Subsequently the properties of persistent currents were studied in Ref. \cite{Buttiker,Gefen1,Gefen2} within different approaches for ballistic and diffusive regimes of conductivity.  It was revealed that the condition of their feasibility consists of the requirement that the size of a system should  be  mesoscopic, i.e. the  radius  of a ring has to be of the order of the electron mean free path at zero temperature.

Another insight into the nature of these currents was obtained after the discovery of the quantum Hall effect in 2D MOSFET structures \cite{Klitzing1980}.
It was shown in Ref.~\cite{MacDonald1984PRB}  that in the rectangular geometry with the characteristic sizes much larger than the electron magnetic length the quantized Hall current may be expressed as the difference between
diamagnetic currents flowing along the two edges
(see also Ref.~\cite{Mineev2007PRB} for a more recent discussion of link between quantum Hall effect and diamagnetism).
When in a state of thermodynamic equilibrium  the chemical potential of 
these edges is the same,  the edge currents cancel each other and the total one caused by the 
applied external magnetic field is zero.

Interestingly, a more simple rectangular geometry was considered in Ref.~\cite{MacDonald1984PRB} two years after
the same problem was studied in the annular geometry \cite{Halperin1982PRB}. This is not surprising because
the annular geometry of the Corbino disk represents a practical realization of the cylinder geometry suggested by Laughlin for
the gedanken experiment explaining quantum Hall effect. The sizes of the disk in Ref. \cite{Halperin1982PRB} are assumed to be macroscopic, i.e. its inner 
and outer radii along  with the width of the ring strongly exceeds the electron magnetic length. Again if the chemical potential of the two edges is the same, the currents at the inner and outer edge flow in the opposite
directions and there is no net current around the annulus.

It could seem that  the contradiction exists between the two above discussed approaches. 
Summing of edge currents in annulus results in their cancellation,  while the one-dimensional treatment of the thin ring 
demonstrates the existence of the persistent current in it. This imaginary controversy 
is related to the reconstruction of the electron spectrum in the annulus as it becomes of the microscopic size. Indeed, when  its dimensions
approach the magnetic length, the staircase of Landau energy levels undergoes 
non-negligible alteration, and  the edge currents start to overlap.

Persistent currents owing to edge effects of the diamagnetic response in the bulk disks and due to the Aharonov–Bohm effect in a mesoscopic samples were initially predicted as the tiny effects, and they were hardly detectable experimentally at those times. Nevertheless, enormous advances in nanotechnology of the last two decades renewed interest to this elusive quantum mechanical phenomenon. Recently the magnetic response of individual gold rings with the typical radii of the order of 1 $\mu m$ and the comparable width indeed has been measured at very low temperatures. It has been found that the response of sufficiently small rings in applied magnetic field can be attributed to the predicted in Ref. \cite{Buttiker, Gefen1, Gefen2} persistent currents. Their amplitudes were found in a rather good agreement to the corresponding theory for quasi-one-dimensional rings \cite{Moler, Glazman2009}.

It is important to note that in all Refs. \cite{Kulik1,Buttiker, Gefen1, Gefen2} the consideration of persistent currents was carried out solely for a quasi-one-dimensional mesoscopic ring within the assumption of its infinitesimal small width. Contrary, in 
Ref.~\cite{Varlamov}  the current density distribution was studied for the macroscopic Corbino disk placed in classically strong magnetic field  under the assumption that all its sizes strongly exceed the electron magnetic length.

The authors of Ref.~\cite{Avishai1993PRB} investigated the persistent current in the same macroscopic annulus being in ballistic regime as a function of electron density. They found the violent fluctuations of the current (in sign and in absolute value) which is quite unusual for systems without disorder. It was demonstrated that these fluctuations result from the overlapping of the sign-changing currents produced by the inner and outer edge states. The same authors in Ref.~\cite{Avishai1993PA}
considered numerically the case of the large disk  with the infinitesimal inner radius in the strong magnetic field.

The goal of the present paper is to establish a bridge between the listed above different approaches for calculation of the  persistent current in the systems of various scales.  In purpose to do this we perform  calculations for the annulus (Corbino disk geometry)  in applied perpendicular homogeneous magnetic field 
with the inner and outer radii arbitrary with respect to the magnetic length and among themselves.
We start from the microscopic solution of the eigenvalues and eigenfunctions problem for the electron in a magnetic field in the case of doubly connected geometry. Basing on it we succeed in the unified way to find the emergence of the first current states in quantum dots, genesis of the persistent currents in quasi-one-dimensional rings, current  oscillatory behavior and the current density profiles in finite size metallic annulus. Finally, we observe fading away of these currents when the sizes of the annulus become substantially macroscopic.

\section{Model and general relations}

\subsection{Model}

We consider an annulus with the inner and the outer radii $r_1$ and $r_2$, correspondingly, subjected to a constant magnetic field $\bf{B}$ applied perpendicularly to its plane described by the Hamiltonian
\begin{equation}
\label{Hamiltonian}
H = \frac{{{{\bf{p}}^2}}}{{2m_e}} = \frac{1}{{2m_e}}{\left( { - i\hbar \bm{\nabla}  + \frac{e}{c}{\bf{A}}\left( {\bf{r}} \right)} \right)^2},
\end{equation}
where $m_e$ is the effective electron mass.
Due to the axial symmetry of the problem it is convenient 
to study it in the polar coordinates, where the vector potential is written in the symmetric gauge ${\bf{A}}\left( {\bf{r}} \right) = (A_r, A_\varphi) =  \frac{1}{2}\left( { 0, Br}\right)$. 
Consequently, the Schr\"{o}dinger equation acquires the form
\begin{widetext}
\begin{eqnarray}
\label{Schrodinger_general}
\left[ { - \frac{{{\hbar ^2}}}{{2m_e}}\left( {\frac{{{\partial ^2}}}{{\partial {r^2}}} + \frac{1}{r}\frac{\partial }{{\partial r}} + \frac{1}{{{r^2}}}\frac{{{\partial ^2}}}{{\partial {\varphi ^2}}}} \right) -
\frac{{i\hbar {\omega _c}}}{2}\frac{\partial }{{\partial \varphi }} + \frac{1}{8}m_e\omega _c^2{r^2}} \right]\psi \left( {r,\varphi } \right) = E\psi \left( {r,\varphi } \right).
\end{eqnarray}
\end{widetext}
Here  $-e<0$ is the charge of the electron and  ${\omega _c} = eB/(m_e c)$ is the cyclotron frequency.
The boundary conditions imposed on the wave function $\psi \left( {{r},\varphi } \right)$ correspond to the impenetrability of the disk edges, i.e. the mandatory requirement 
\begin{equation}
\psi \left( {{r_1},\varphi } \right) = \psi \left( {{r_2},\varphi } \right) = 0.
\label{imp}
\end{equation} 
Separation of the variables in Eq. (\ref{Schrodinger_general})
\begin{equation}
\label{solution_general}
\psi \left( {r,\varphi } \right) = f\left( r \right){e^{-im\varphi }}
\end{equation}
reduces it to the differential equation for the radial component of the wave function
\begin{equation}
\label{equation_radial}
\left[ {\! -\!\frac{{{\hbar ^2}}}{{2m_e}}\left( {\frac{{{\partial ^2}}}{{\partial {r^2}}}\! + \!\frac{1}{r}\frac{\partial }{{\partial r}} \!-\!\frac{{{m^2}}}{{{r^2}}}} \right) + \frac{1}{8}m_e\omega _c^2{r^2}} \right]f\left( r \right)\! = \!\tilde Ef\left( r \right),
\end{equation}
where the energy $\tilde E$  is shifted in respect to $E$  as  $\tilde E = E + \hbar \omega _c m/2$.
Introducing the dimensionless energies  
$\varepsilon  = {E}/(\hbar {\omega _c})$, 
$\tilde\varepsilon = {\tilde E}/( {\hbar {\omega _c}})$ 
and the dimensionless variable $\rho  = r/l$ (where $l = 
\sqrt {\hbar c/(|e| B)}$
is the magnetic length) one can simplify 
Eq.~(\ref{equation_radial}):
\begin{equation}
\label{equation_radial_new}
\left[ {\frac{{{\partial ^2}}}{{\partial {\rho ^2}}} + \frac{1}{\rho }\frac{\partial }{{\partial \rho }} - \frac{{{\rho ^2}}}{4} + 2\tilde\varepsilon   - \frac{{{m^2}}}{{{\rho ^2}}}} \right]f\left( \rho \right) = 0.
\end{equation}
Its general solution can be written in terms of two Whittaker functions ${M_{\kappa ,\mu }}\left( z \right)$  and  ${W_{\kappa ,\mu }}\left( z \right)$:
\begin{equation}
\label{solution_radial}
f\left( \rho \right) = \frac{1}{\rho }\left[ {{C_1}{M_{\tilde\varepsilon  ,\frac{{\left| m \right|}}{2}}}\left( \frac{\rho ^2}{2} \right) + {C_2}{W_{\tilde\varepsilon  ,\frac{{\left| m \right|}}{2}}}\left( \frac{\rho ^2}{2} \right)} \right].
\end{equation}
Below we will analyze the spectral properties of Eq. (\ref{equation_radial_new}) and the asymptotic behavior of radial functions (\ref{solution_radial}).

\subsection{Dispersive Landau levels in the Corbino geometry}

Applying of the  boundary conditions (\ref{imp}) at Eq. (\ref{solution_radial}) yields a transcendental equation for the energy levels $\tilde\varepsilon_{n,m} \equiv \tilde\varepsilon$:
\begin{equation}
\label{energy_levels}
{W_{\tilde\varepsilon  ,\frac{{\left| m \right|}}{2}}}\left( {\frac{\rho _1^2}{2}} \right)\!{M_{\tilde\varepsilon  ,\frac{{\left| m \right|}}{2}}}\left( {\frac{\rho _2^2}{2}} \right)\! -\!{M_{\tilde\varepsilon  ,\frac{{\left| m \right|}}{2}}}\left( {\frac{\rho _1^2}{2}} \right)\!{W_{\tilde\varepsilon  ,\frac{{\left| m \right|}}{2}}}\left( {\frac{\rho _2^2}{2}} \right) \!= \!0.
\end{equation}
Here the two quantum numbers appear, viz. the first (principal) $n$  
corresponds to the level number in Landau  problem, while the second, azimuthal one $m$, characterizes the angular momentum and the latter is an analogue of the wave-vector component $k_y$ in the Landau gauge \cite{Landau}.  Let us recall, that in the case of the Landau gauge the quantum number $k_y$ determines the position of the potential minimum $x_0=l^2k_y$ in the coordinate space. In the case under consideration its role passes to the value $r_m=\sqrt{2|m|}l$, which is nothing else as the position along the radial coordinate of the maximum in the probability of the electron state with quantum number $m$ for the given $n$.

The same impenetrability boundary conditions allow to simplify  the form of radial part of the wave function (\ref{solution_radial})
\begin{eqnarray}
f_{nm}\left( \rho \right) = \frac{C}{\rho }\left [W_{{\tilde\varepsilon  ,\frac{{\left| m \right|}}{2}}}\left( {\frac{\rho _1^2}{2}} \right)\!{M_{\tilde\varepsilon  ,\frac{{\left| m \right|}}{2}}}\left( {\frac{\rho^2}{2}} \right)\right. -\nonumber \\
\left. {M_{\tilde\varepsilon  ,\frac{{\left| m \right|}}{2}}}\left( {\frac{\rho _1^2}{2}} \right)\!{W_{\tilde\varepsilon  ,\frac{{\left| m \right|}}{2}}}\left( {\frac{\rho^2}{2}} \right) \right],\label{psi_func}
\end{eqnarray}
leaving in it the only constant $C$. The latter is determined from the 
normalization condition:
\begin{equation}
\label{constant}
\begin{split}
{(C l)^{-2}}=2\pi & \int\limits_{{\rho _1}}^{{\rho _2}} {\frac{d \rho}{\rho }}  \left[  {{W_{\tilde \varepsilon ,\frac{{\left| m \right|}}{2}}}\left( {\frac{\rho _1^2}{2}} \right)
{M_{\tilde \varepsilon ,\frac{{\left| m \right|}}{2}}}
\left( {\frac{\rho ^2}{2}} \right)-} \right.  \\
& {\left. {{M_{\tilde \varepsilon ,\frac{{\left| m \right|}}{2}}}
\left( {\frac{\rho _1^2}{2}} \right){W_{\tilde \varepsilon ,\frac{{\left| m \right|}}{2}}}\left( {\frac{\rho^2}{2}} \right)} \right]^2}.
\end{split}
\end{equation}

Accordingly, the eigenfunction of the one particle problem is 
\begin{equation}
\label{psi_nm}
\psi _{nm}\left( \mathbf{r}  \right) = f_{nm}\left( r  \right) e^{- i m \varphi} 
\end{equation}
(see Eqs.~(\ref{solution_general}) and (\ref{psi_func})).

Let us note that the Whittaker functions in Eq.~(\ref{energy_levels}), in the case when the parameter $\tilde \varepsilon_{n,m}$ equals to
\begin{equation}
\label{parameter_trivial}
\tilde \varepsilon_{n,m}  = n + \frac{1}{2}\left( {\left| m \right| + 1} \right), \qquad n=0,1, \ldots,
\end{equation}
turn out linearly dependent and they reduce to Laguerre polynomials. It is why these trivial solutions, corresponding to the infinite system with the spectrum (this problem in the symmetric gauge was addressed in \cite{Frenkel1930})
\begin{equation}
\label{energy_levels_trivial}
{\varepsilon _{n,m}} = n + \frac{1}{2}\left( {\left| m \right| - m + 1} \right), \quad n=0,1, \ldots, \quad m \geq -n,
\end{equation}
we exclude from consideration.

\begin{figure}
\includegraphics[width=0.49\columnwidth]{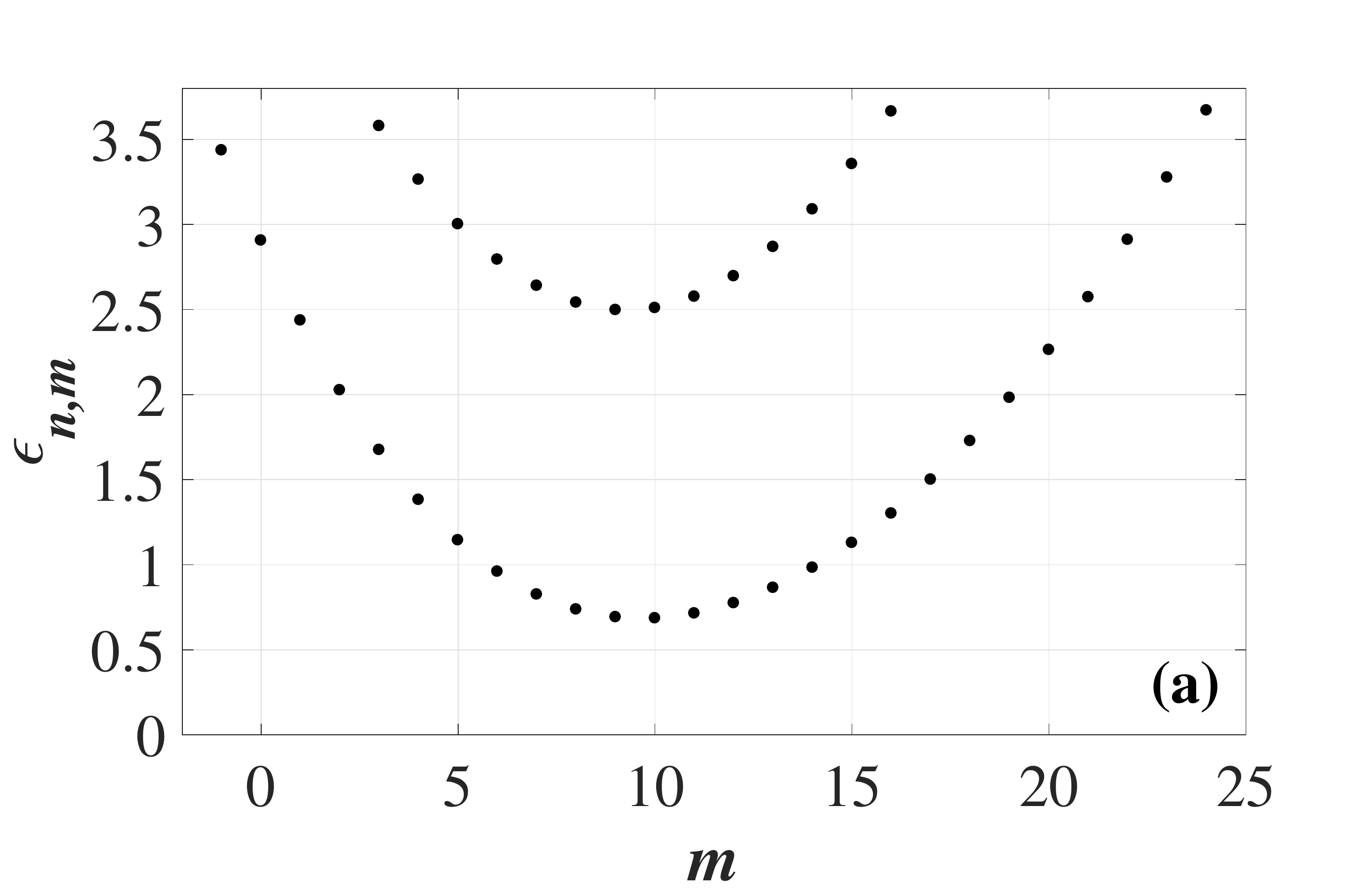}
\includegraphics[width=0.49\columnwidth]{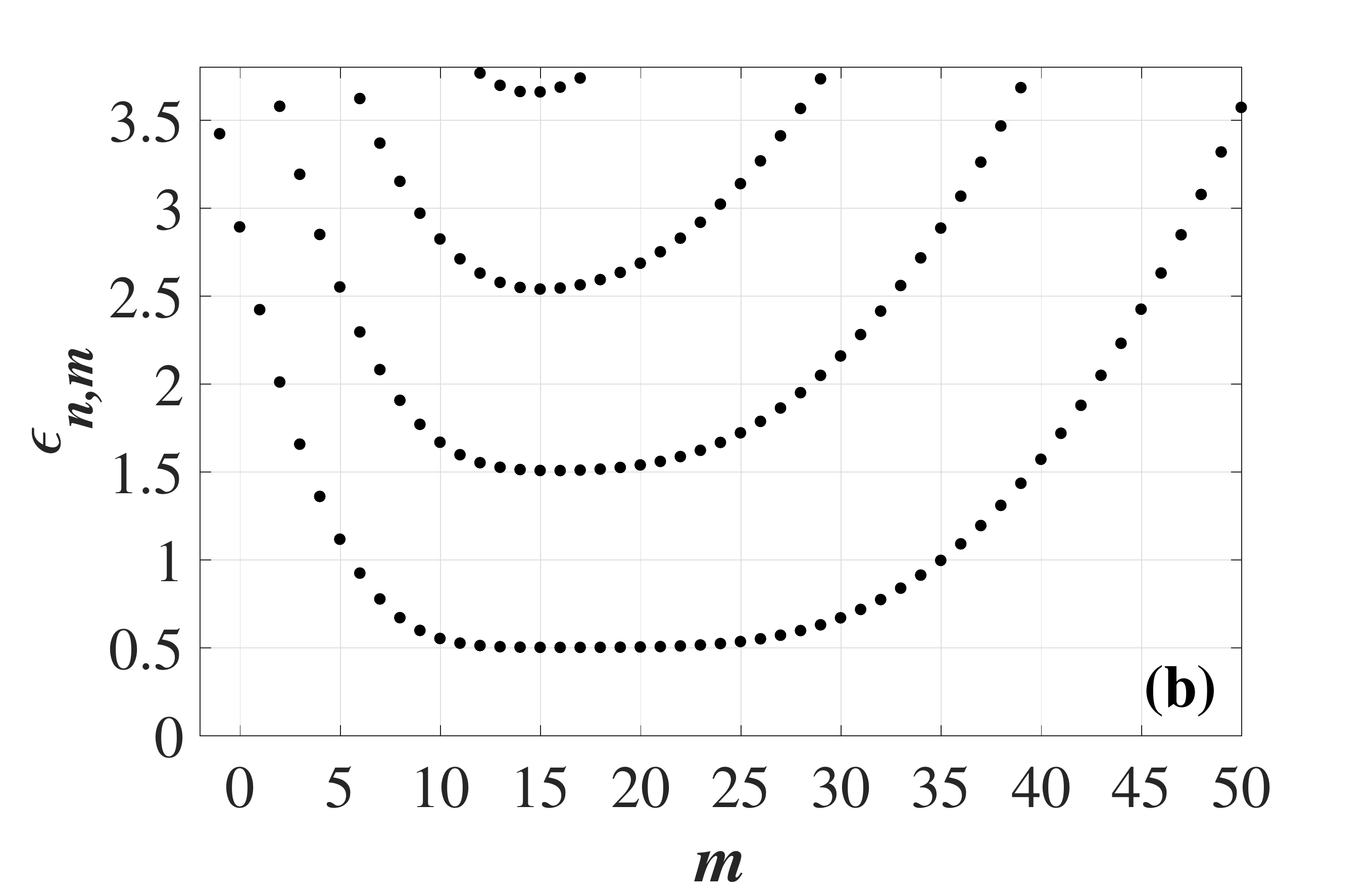}
\includegraphics[width=0.49\columnwidth]{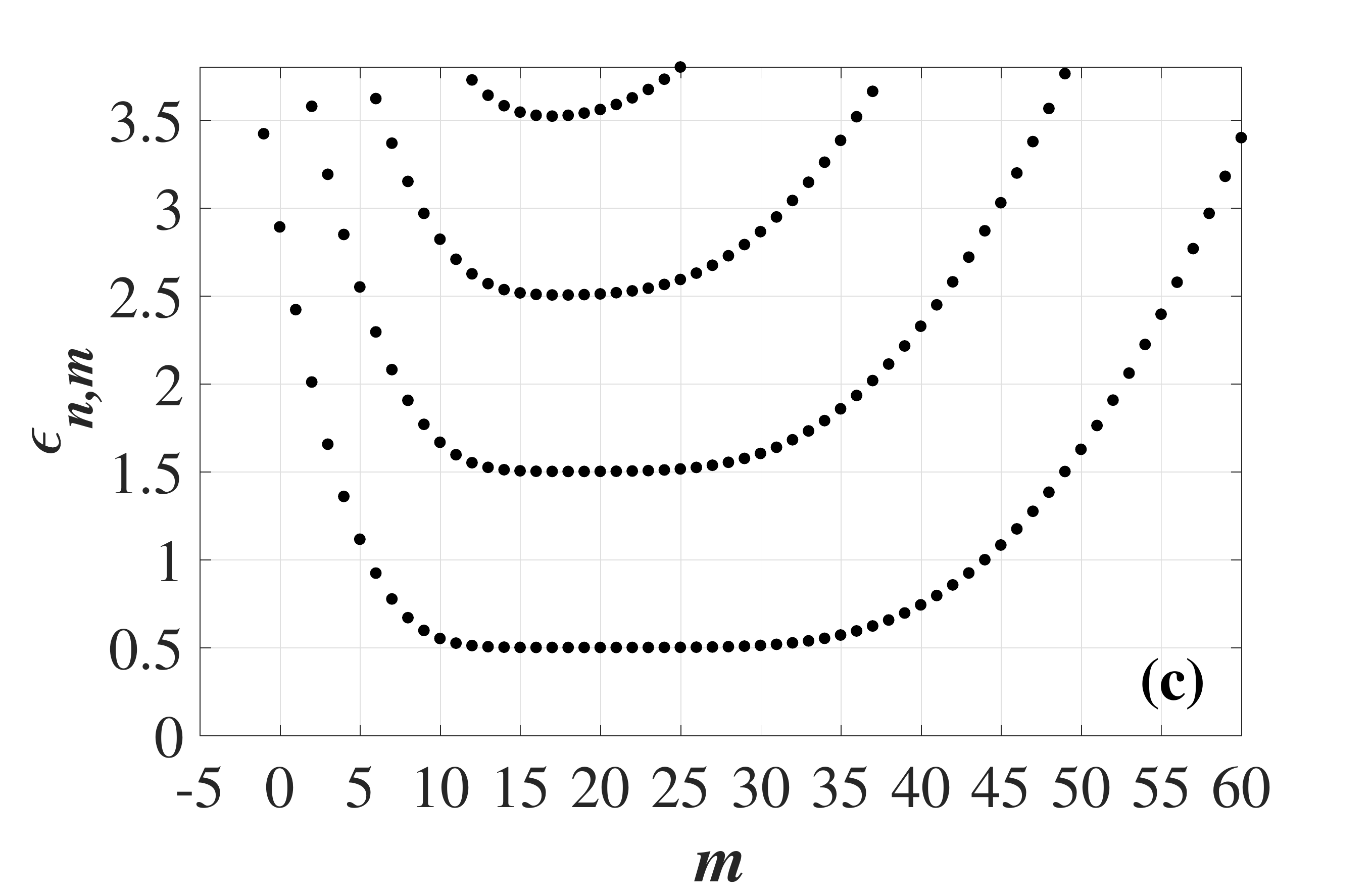}
\includegraphics[width=0.49\columnwidth]{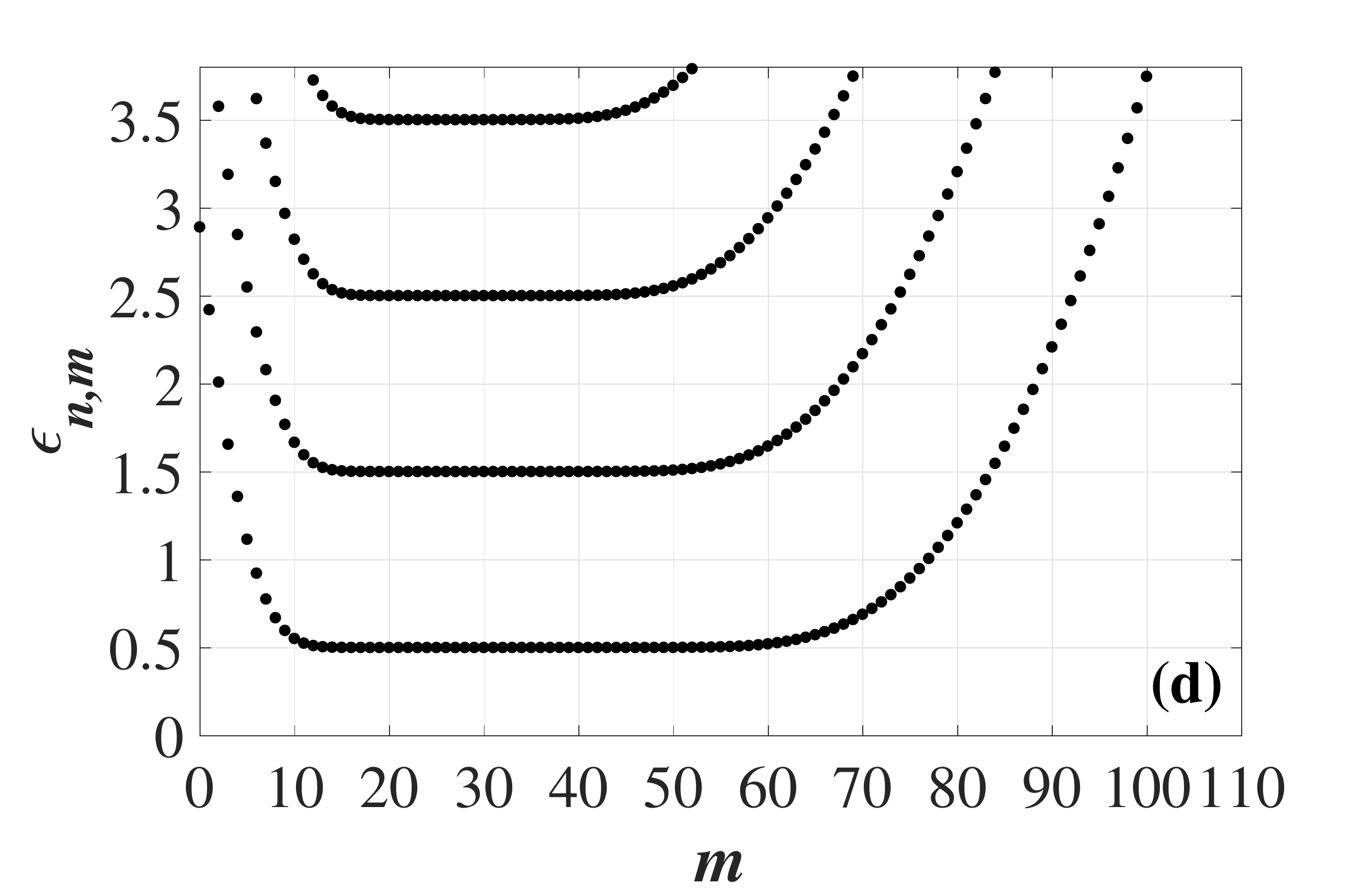}
\caption {Energy levels as a function of angular quantum number $m$ for an annulus  with a fixed inner radius $\rho_1 = 3$ and different outer radii $\rho_2 =6$ (a), $\rho_2 =9$ (b), $\rho_2 =10$ (c) and  $\rho_2 =13$ (d).}
\label{enery_level_general}
\end{figure}

In Figure~\ref{enery_level_general} one can see the series of the energy levels $\varepsilon_{n,m}$ as the function of the angular
quantum number $m$ for different inner and outer radius of the disk obtained from the numerical solution of Eq.~(\ref{energy_levels}). 
One observes that as the outer radius $\rho_2$ increases the Landau levels inside the ring flatten and approach the values (\ref{energy_levels_trivial}) for the 
infinite system.

\subsection{Persistent currents}

\subsubsection{General expressions}

The current density operator in the representation of field operator $\psi$ is 
\begin{equation}
\label{current_general}
{\bf{\hat j}} =   \frac{{ie\hbar }}{{2m_e}}\left( {{\psi^ * }\bm{\nabla} \psi  -  
\bm{\nabla} {\psi^ * }} \psi \right) - \frac{e^2}{{m_e c}}{\bf{A}}\psi^* {\psi }.
\end{equation}
For the Fermi gas with chemical potential $\mu$
the  current density  expectation value reads \cite{AGD}
\begin{equation}
\label{current_density}
\begin{split}
& \mathbf{j} \left( \mathbf{r} \right) =  - \frac{e}{{2m_e}}
\left(  \pmb{\pi} \left( \mathbf{r} \right) + { \pmb{\pi}^* }
\left( \mathbf{r}^\prime \right) \right) \times \\
& \sum_{\substack{m = -\infty\\ n=0 } }^\infty  
{{\left.  {\psi _{nm}}\left( \mathbf{r}  \right)
\psi_{nm}^*\left( \mathbf{r}^\prime  \right) \right|}_{\mathbf{r}^\prime = \mathbf{r}  }}n_{F}\left( E_{n,m}  \right),
\end{split}
\end{equation}
where $\pmb{\pi} = -  i \hbar  \bm{\nabla}  + \frac{e}{c} \mathbf{A}(\mathbf{r})$ is the gauge invariant momentum operator, $\psi_{nm}$ is given by Eq.~(\ref{psi_nm}),
and $n_F (E)=
[\exp ((E - \mu) /T) + 1]^{-1}$
is the Fermi-Dirac distribution function. In the considered limit of low temperatures ($T \ll \hbar \omega_c$) is reduced to the Heaviside function $\theta(\mu-E_{n,m})$. 

Generally speaking the chemical potential $\mu$ is the sophisticated function of the electron density and applied magnetic field (see for example \cite{Mineev1}). 
The electron density is fixed in a closed system.  Moreover, in
the following we will be interested in the energy level dispersion and
current distribution as the functions of the system size for a chosen value of the 
magnetic field. Thus in the further consideration we assume chemical potential 
as the constant.

Based on the expressions for radial and tangential components of a momentum
\begin{equation}
\label{momentum}
{\pi_r} =  - i\hbar \frac{\partial }{{\partial r}}, \qquad
{\pi_\varphi } =  - \frac{{i\hbar }}{r}\frac{\partial }{{\partial \varphi }} + \frac{{eBr}}{{2c}},
\end{equation}
one can find the corresponding expressions for the current density:
\begin{equation}
\label{radial_current}
\begin{split}
{j_r} =  - \frac{ie \hbar}{2m_e}
& \left( \frac{\partial}{\partial r} - \frac{\partial}{\partial r^\prime} \right)
  \sum_{\substack{m = -\infty\\ n=0 } }^\infty   \theta(\mu- E_{n,m}) \times \\
 & \left. \psi_{nm} \left( {r,\varphi } \right) \psi_{nm}^* 
 \left( r^\prime,\varphi  \right) \right|_{r^\prime = r} \equiv 0  
\end{split}
\end{equation}
and
\begin{equation}
\label{current_density_up1}
\begin{split}
{j_\varphi }(r) = \frac{{e\hbar }}{{2m_e{l^2}}}\sum_{\substack{m = -\infty\\ n=0 } }^\infty {\left( {\frac{{r_m^2}}{r}{\rm{sgn}}(m) - r} \right)}  \times \\
{\left| {{\psi _{nm}}\left( {r,\varphi } \right)} \right|^2}\theta(\mu-E_{n,m}).
\end{split}
\end{equation}

Full current flowing in the disk is
\begin{equation}
\label{full_current}
I = \int_{r_1}^{r_2}j_\varphi (r) dr =\sum_{\substack{m = -\infty\\ n=0 } }^\infty I_{nm} \theta(\mu - E_{n,m})
\end{equation}
with the partial component $I_{nm}$ carried by the state with definite quantum numbers $n,m$:
\begin{equation}
\label{I_nm}
\begin{split}
I_{nm} & =   \frac{ e \hbar}{2m_e}  {\int\limits_{{r_1}}^{{r_2}} { \left( {  \frac{{2m}}{r } - \frac{r}{l^2} } \right)} } f_{nm}^2\left( r \right) d r \\
& =\frac{ e \hbar}{2m_e} \left( 
2m\int\limits_{{r_1}}^{{r_2}} 
\frac{{d r}}{r } f_{nm}^2\left( r \right) 
- \frac{1}{2 \pi l^2} \right),
\end{split}
\end{equation}
where in the second term we used the normalization condition
\begin{equation}
\label{normalization-f}
\int d^2 r |\psi_{nm} (r, \varphi)|^2 =
2 \pi \int_{r_1}^{r_2} r d r f_{nm}^2(r) =1.
\end{equation}
One can see that 
Eqs.~(\ref{full_current}) and (\ref{I_nm}) are in agreement to Eq.~(7) in 
\cite{Halperin1982PRB}. 
Note that the first paramagnetic part in Eqs.~(\ref{current_density_up1}) and (\ref{I_nm})
originates from the gradient term in  Eq. (\ref{current_general}) and  is related to the spacial inhomogeneity of the current flow in the disk. 
The second term of the corresponding equations  $ \sim B$ is diamagnetic.

\subsubsection{Byers-Yang formula}
\label{sec:BY-formula}

It is worth to mention that a more complicated problem with the annulus placed in the constant magnetic field and threaded by the flux $\Phi$ can still be considered basing on the solution  (\ref{solution_radial}) in
terms of the Whittaker functions. This occurs because adding a vector potential  
${\bf{A}}_\eta\left( {\bf{r}} \right) =  \left( { 0, \Phi_0 \eta/(2 \pi r)}\right)$
corresponding to the magnetic field 
$\mathbf{B}_\eta (\mathbf{r}) = \nabla \times {\mathbf{A}}_\eta 
= \mathbf{e}_z \Phi_0 \eta \delta^2 (\mathbf{r})$ does not alter the structure of Eq.~(\ref{Schrodinger_general}).
One can easily check that the corresponding
solution for the problem that involves a superposition of the constant field and flux can be written by mere replacement $m \to 
m - \eta$. The energy spectrum of the infinite system is still given  by
Eq.~(\ref{energy_levels}) with the shifted 
azimuthal quantum number.

Under certain conditions the current $I_{nm}$ carried by the state with definite
quantum numbers $n,m$ can be found using the Byers-Yang formula \cite{Byers1961PRL,Imry.book}: 
\begin{equation}
\label{Byers-Yang}  
I_{nm} = - \frac{e}{h} \frac{\partial E_{n,m}(\eta)}{\partial \eta} = \frac{e}{h} \frac{\partial E_{n,m}(\eta)}{\partial m}.  
\end{equation}
The latter expresses  the fact that the persistent current is the thermodynamic quantity
conjugated to the flux through the ring.
Although initially the Byers-Yang formula was introduced for a system with a hole threaded
by the flux, the second equality in Eq.~(\ref{Byers-Yang})
(see Ref.~\cite{Avishai1993PA} for a detailed discussion)
allows one to use it for the description of the current in the
system with hole in a constant magnetic field with $\eta=0$. Yet, considering the spectrum for the infinite system, Eq.~(\ref{energy_levels}),
one can see that this formula is ill defined for $m=0$. We note that the Byers-Yang formula follows from the Hamiltonian given by  Eq.~(\ref{Hamiltonian}) if one utilizes the Hellman-Feynman theorem. We will use this formula below to estimate the values of $I_{nm}$.

Finally we note that in what follows we use the dimensionless units for the brevity of notations. Yet we will restore the units to underline the physics.

\section{The cases amenable to analytical solution}

\subsection{Asymptotic analysis of the eigenvalue problem for wide annulus}
\label{sec:Corbino-wide}

\begin{figure}
\includegraphics[width=0.49\columnwidth]{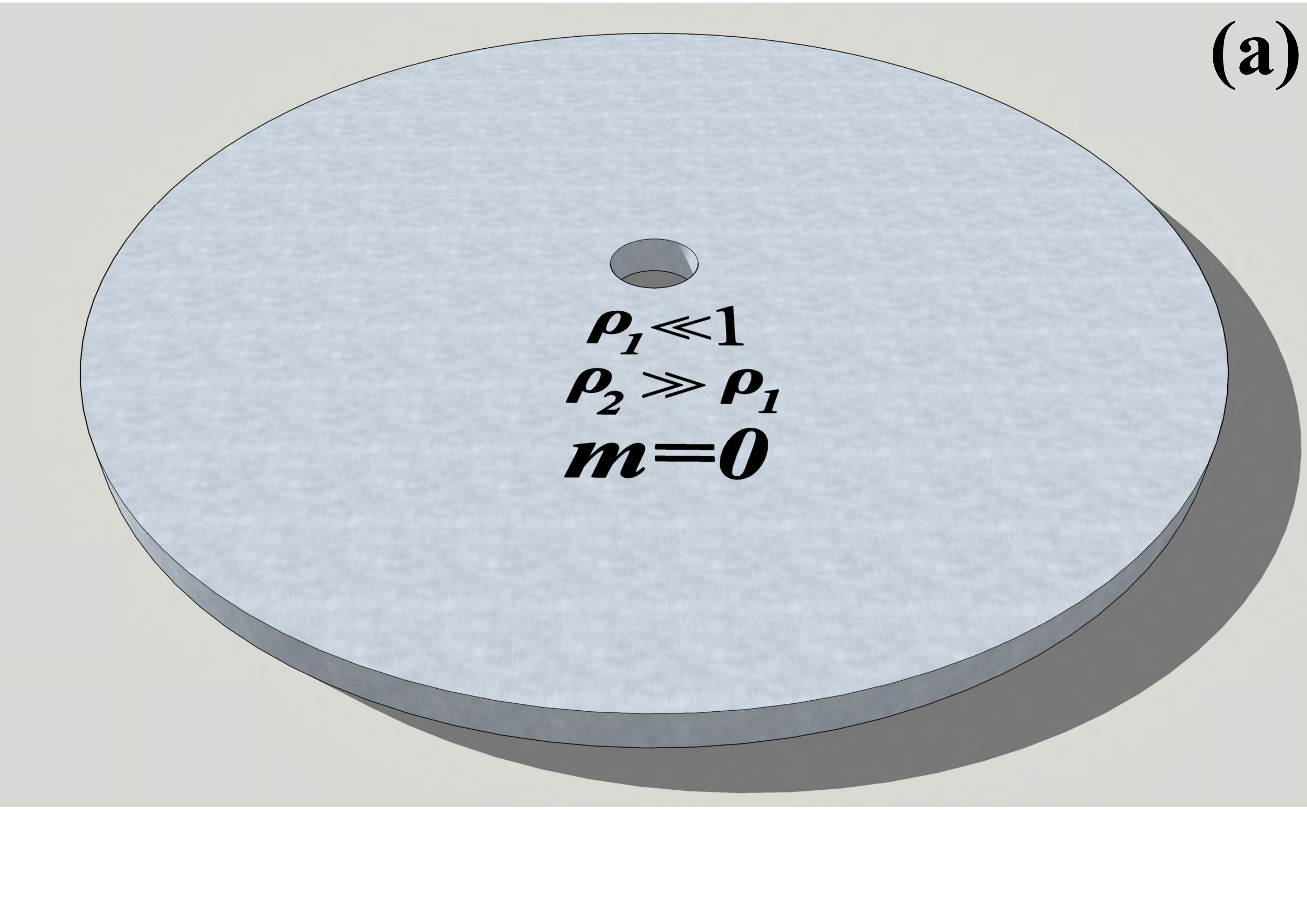}
\includegraphics[width=0.49\columnwidth]{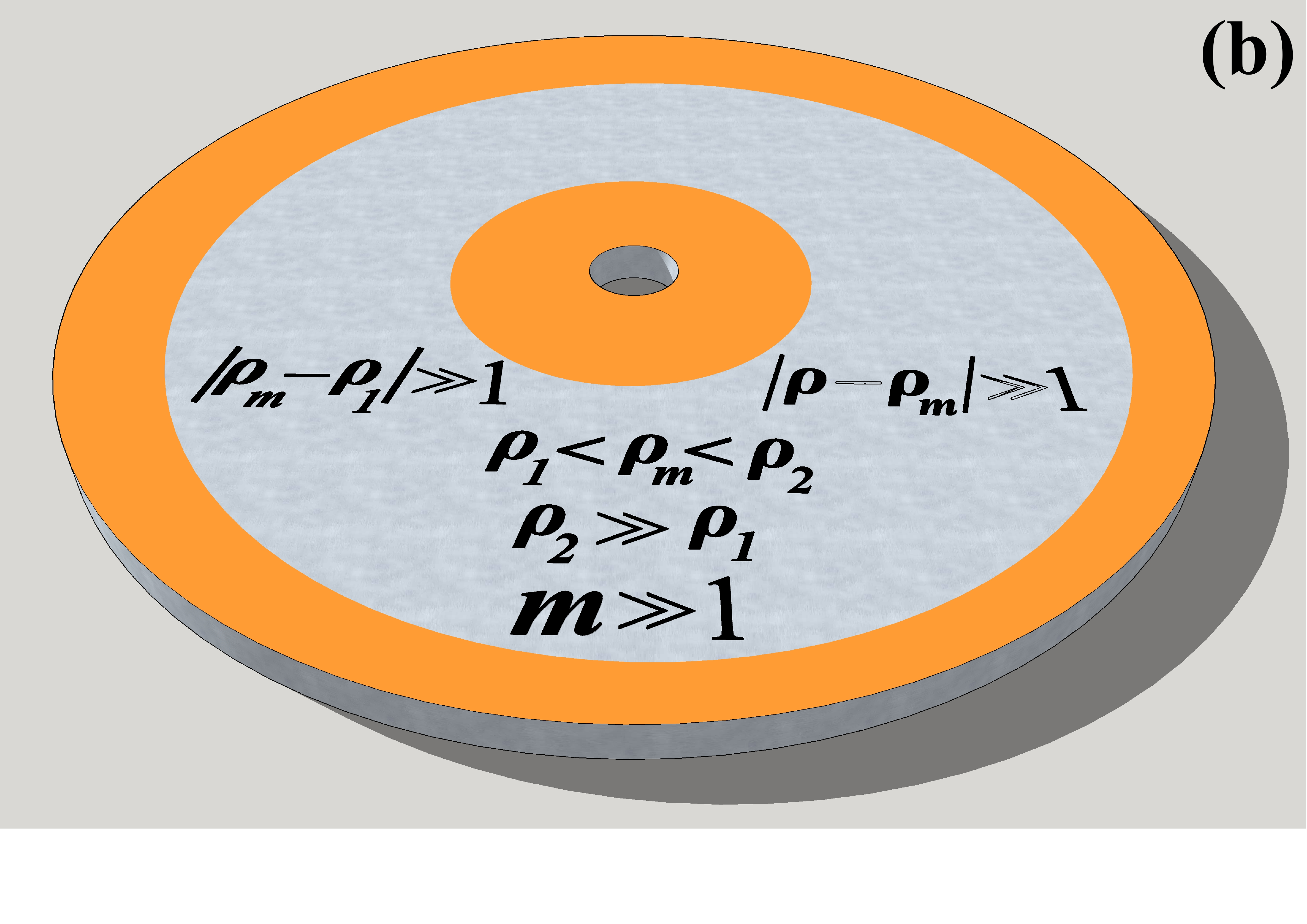}
\includegraphics[width=0.49\columnwidth]{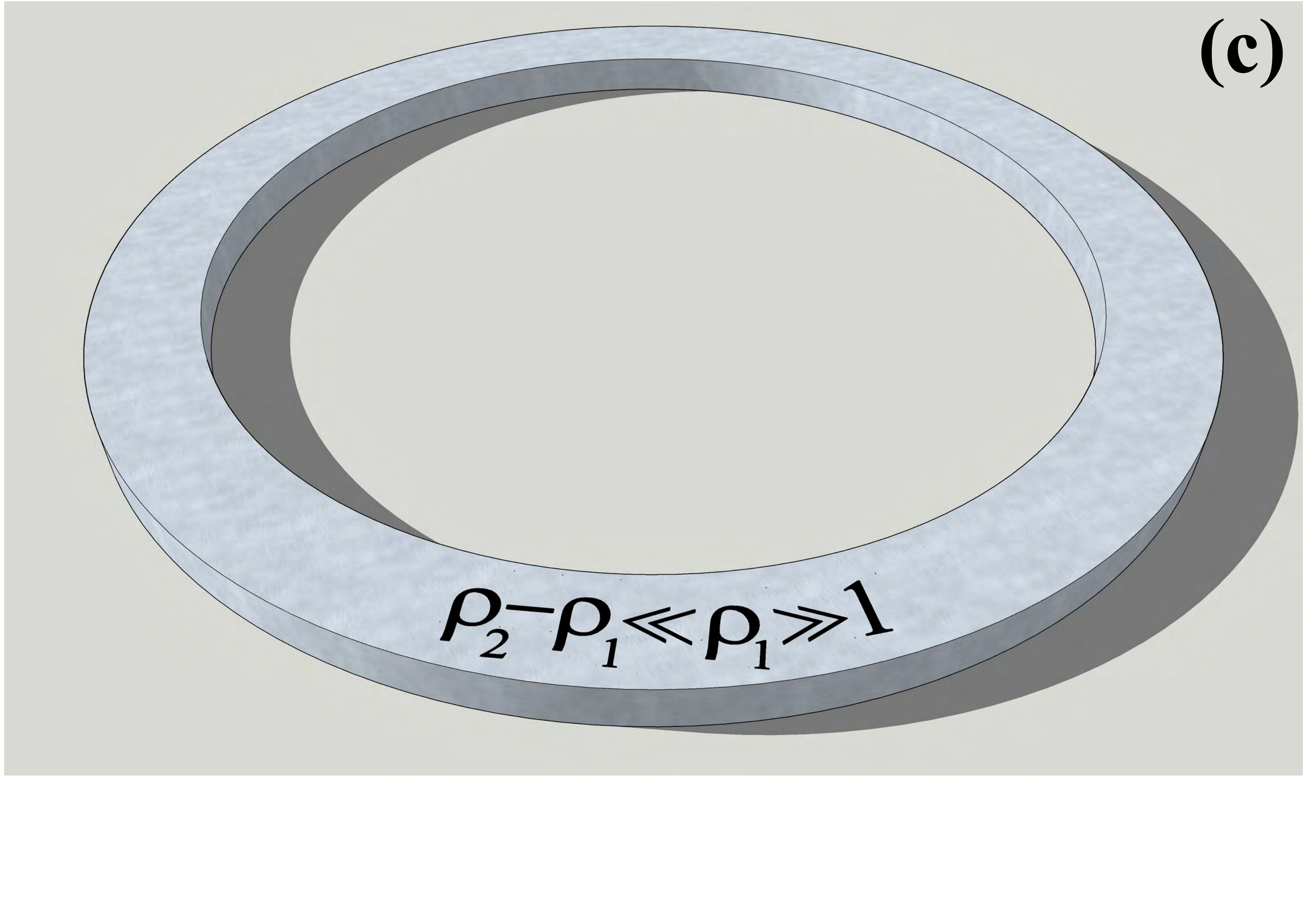}
\includegraphics[width=0.49\columnwidth]{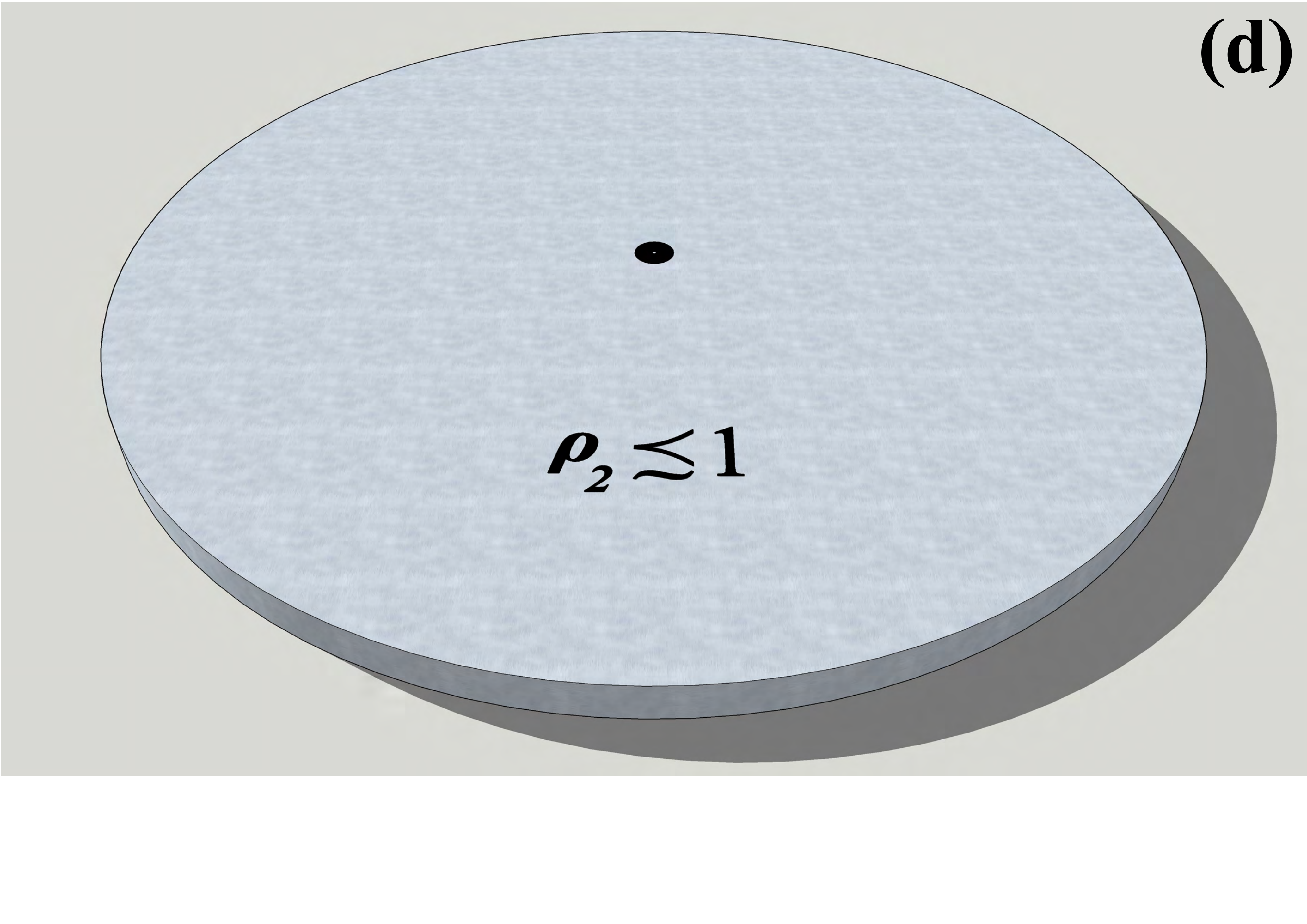}
\caption{Graphical representation of applicability of the analytical expressions for energy levels in a annulus under the conditions specified in figures (see details in the text). (a) Eq. (\ref{energy_levels_m=0_new}) is valid in the entire disk. (b)  Eq. (\ref{energy_levels_k=inf_bulk}) is for the bulk region, while inner and outer orange domains correspond Eq. (\ref{energy_levels_k=inf_r1}) and 
Eq.~(\ref{energy_levels_k=inf_r2}) respectively. (c)  Eq. (\ref{narrow-spectrum-units}) corresponds for a narrow annulus (ring). (d) Eq. (\ref{energy_levels_small_disc_new}) is applicable for the small annulus with the infinitesimal inner radius. The proportions are not to scale. The figures are illustrative in nature.}
\label{sketch_energy_sol}
\end{figure}

The solution of transcendental equation (\ref{energy_levels}) admits the rich variety of asymptotic representations. Below we analyze the approximation of a wide annulus $\rho_2 \gg \rho_1$, and, besides the first case with  $m =0$, in the following the large angular momentum limit ($m \gg 1$)  will be in the focus of our discussion. 
For the sake of convenience they are graphically classified in Figure~\ref{sketch_energy_sol}.

Another important parameter is the cyclotron radius of the orbits for the most essential electrons being at the highest Landau levels (which energies are close to the chemical potential): $r_c=mv_F/ (eB)=l \sqrt{2\mu/ (\hbar \omega_c)}$. It determines the effective width of the region where the edge currents flow and in the following consideration will remain arbitrary with respect to the inner and outer radii of the annulus.

\subsubsection{The states with zero angular momentum in a disk with a small hole}
\label{sec:m=0}

 Let us start from consideration of the electron spectrum in a wide disk ($\rho_2 \gg 1$) with the small hole ($\rho_1 \ll 1 $).
In this case one can find the explicit expression for the energy levels corresponding the states with  $m=0$. Basing on the asymptotic expressions of the Whittaker functions with $m=0$ for small arguments one can arrive to Eq. (\ref{energy_levels_k=0}) (see details of the derivation in Appendix \ref{sec:A} in parts 2 and 3). Corresponding energy levels for not very large quantum numbers $n+1 \ll 1/\rho_1^2$ are given by:
\begin{equation}
\label{energy_levels_m=0_new}
\varepsilon_{n,0}  = n + \frac{1}{2} + \frac{1}{{\ln \left( {\frac{2}{{\rho _1^2}}} \right) - \psi \left( {n + 1} \right) - 2\gamma }}, 
\end{equation}
where $\psi(z)$ is the digamma function and $\gamma =0.577...$ is the Euler-Mascheroni constant. One can see, that  the Landau equidistant ``ladder'' distorts, and this distortion increases with the growth of the level number.

\subsubsection{The states with large angular momentum close to the edges}
\label{sec:large-m}

In an annulus with the radius of the inner hole larger than magnetic length ($ \rho_1  \gtrsim  1$) the edge states are formed in the both border areas. Near the edges 
when the center of the wave function $\rho_m=\sqrt{2|m|} $
satisfies the condition $|\rho_m - \rho_i| < 1$ ($i=1,2$)
the energy levels have the following form (see Eq.~(\ref{energies_hole_bulk_2}) in the Appendix~\ref{sec:A})
\begin{equation}
\label{energy_levels_k=inf_r1}
\begin{split}
\varepsilon_{n,m}&= 2 \left (n+\frac{3}{4} \right)-\frac{\Gamma(n+1/2)}{\pi n!}(2n+1)\frac{\rho_m^2 -\rho_1^2}{\rho_m}  \\
& + 2 G(2n+2) \left(\frac{\Gamma(n+3/2)}{\pi n!}\right)^2\frac{(\rho_m^2-\rho_1^2)^2}{\rho_m^2},
\end{split}
\end{equation}
\begin{equation}
\label{energy_levels_k=inf_r2}
\begin{split}
\varepsilon_{n,m}&= 2 \left(n+ \frac{3}{4} \right)-\frac{\Gamma(n+1/2)}{\pi n!}(2n+1)\frac{\rho_2^2-\rho_m^2}{\rho_m}  \\
& +2 G(2n+2) \left(\frac{\Gamma(n+3/2)}{\pi n!}\right)^2\frac{(\rho_2^2-\rho_m^2)^2}{\rho_m^2},
\end{split}
\end{equation}
respectively, with $G(2z) = \psi(z+1/2) - \psi(z)$.
They correspond to the energies of the skipping electrons subjected to the parabolic potential with the minimum shifted  with respect to the inner/outer edge. 
In the first term of these expressions one can recognize the spectrum for the harmonic oscillator with reflecting wall at the minimum of potential 
(see the Problem 2.12 in Ref.~\cite{Galitsky_problems}). 
These peculiarities  of the electron energy levels near the edges of the ribbon placed in magnetic field  were  anticipated in \cite{Halperin1982PRB} (also cf. Eqs.~(7) and (9) in Refs.~\cite{Heuser,Varlamov}, respectively).

\subsubsection{The states with large angular momentum far from the edges}

In the bulk of the disk not too close to the edges, when $\rho_1<\rho_m< \rho_2$ and 
$|\rho_m-\rho_i|\gg 1$, 
the electron states are localized. The corresponding 
spectrum tends to the Landau one (see details of the derivation in Appendix~\ref{sec:A} in part 5):
\begin{equation}
\label{energy_levels_k=inf_bulk}
\begin{split}
& \varepsilon^{(i)}_{n,m}=n+\frac{1}{2}+\frac{1}{\sqrt{2\pi}n!}x_i^{2n+1}e^{-x_i^2/2},  \\
& x_i^2=\rho_m^2(\rho_i^2-1-\ln\frac{\rho_i^2}{\rho_m^2}),
\quad \rho_m = \frac{r_m}{l},
\end{split}
\end{equation}
and corresponds to the energy spectrum of the ribbon in magnetic field \cite{Heuser}.

\subsubsection{Currents carried by the states with large angular momentum}

To find the current $I_{nm}$ carried  by  the  state  with definite quantum 
numbers $n,m$ in the different regions of the annulus
we apply Byers-Yang formula (\ref{Byers-Yang}).

First, it is easy to see from Eq.~(\ref{energy_levels_k=inf_bulk}) that inside the disk,
$\rho_1 < \rho_m < \rho_2$ and $|\rho_m - \rho_{i}| \gg 1 $  the current $I_{nm} =0$, because
the Landau levels are flat up to the exponentially small correction in the momentum $m$.

Near the edges of the disk, $|\rho_m - \rho_i | < 1$,
from Eqs.~(\ref{energy_levels_k=inf_r1}) and (\ref{energy_levels_k=inf_r2})
one obtains the currents

\begin{equation}
I_{n}(\rho_i) \equiv I_{nm}(\rho_i) = (-1)^{i} \frac{\Gamma(n+1/2) (2n+1)}{\pi^2 n!} \frac{e \omega_c}{\rho_i},   
\end{equation}
where we set $\rho_m = \rho_i$, so that the partial contributions $I_{nm}(\rho_i)$ are independent on $m$. Let us note that the currents at the inner and outer edges of the disk do  not coincide due to the difference in curvatures. 

The partial contribution to the full persistent current is
\begin{equation}
\label{Inmlast}
I_{n}^{\mathrm{tot}} = - {\mathcal J}_0\frac{2 \Gamma(n+1/2)  (2n+1)}{\pi n!}  
\frac{a_0^2}{l} \left(\frac{1}{r_1}-\frac{1}{r_2} \right),  
\end{equation}
where in order to explicitly highlight the current dimensionality we used the value of the current carried 
by an electron in hydrogen atom
\begin{equation}
{\mathcal J}_0 = \frac{m_e e^5}{2 \pi \hbar^3} \approx \SI{1.05e-3}{A} 
\end{equation}
with and $a_0 = \hbar^2/(m_e e^2)  \approx \SI{0.053}{nm} $ as the Bohr radius. For future comparison with experiment it worth to note that the magnetic length $l= \SI{26}{nm}/\sqrt{B[\mbox{T}]}$ and
the field is measured in Tesla. One can see that in the simple rectangular geometry when 
$r_1, r_2 \rightarrow \infty$ the current (\ref{Inmlast}) turns zero \cite{Heuser}. 

Let us consider the effect of discussed persistent current on the quantum Hall effect measurements in the macroscopic ($2 r_1 =\SI{0.9}{mm} $ and $2 r_2 =\SI{3.9}{mm}  $) annulus \cite{Dolgopolov1992}. 
Assuming that the potential difference between the disk edges is kept to be zero, we estimate the magnitude of the current
$I_{n}^{\mathrm{tot}}$ induced by the disk curvature. Since the filling factor in  Ref.~\cite{Dolgopolov1992} was below 3, we restrict our evaluation only by
one channel, what gives 
$|I_{n}^{\mathrm{tot}}|  = \SI{6.92e-13}{A}$ for $B = \SI{10}{T}$ and $n=0$.  
Such small additional persistent current cannot 
affect the measurements of the quantum Hall effect (for example, in Ref. \cite{Dolgopolov1992} corresponding currents were  five orders more: $\sim \SI{e-8}{A}$).

In order to estimate the magnitude of the total persistent current in such a ring basing on Eq.~(\ref{Inmlast}) it is necessary to perform in it the summation over the quantum numbers $n,m$. The principal quantum number $n$ changes over all filled states below the chemical potential up to its maximal value $ \mu/{(\hbar \omega_c)} \gg 1$, while the azimuthal quantum number $m$ can be fixed by the value of radius $r_2 = \sqrt{2|m|}l$.  Summation in Eq.~(\ref{Inmlast}) is easily performed using the Stirling's formula what results in:
\begin{equation}
\label{Inmlast_update}
\begin{split}
I^{\mathrm{tot}} & = - \frac{8{\mathcal J}_0}{3 \pi}  \left( \frac{\mu}{\hbar \omega_c} \right)^{3/2}
\frac{a_0^2}{l} \left(\frac{1}{r_1}-\frac{1}{r_2} \right) \\ 
& = - \frac{2 \sqrt{2}{\mathcal J}_0}{3 \pi} \frac{a_0^2 r_c^3}{l^4} \left(\frac{1}{r_1}-\frac{1}{r_2} \right).
\end{split}
\end{equation}

Bearing in mind the above use of the Byers-Yang formula one has to remember that Eq.~(\ref{Inmlast_update}) is valid only when the cyclotron radius $r_c$ is much smaller than the width
of the disk, $r_c \ll r_2-r_1$, when the edge currents do not overlap.
One can see that the persistent current  Eq.~(\ref{Inmlast_update}) is inversely proportional to magnetic field and is determined by the difference of the edge curvatures.

\subsection{The spectral problem and persistent currents in a narrow annulus}

\subsubsection{Energy spectrum}

For an annulus with the large inner radius as it becomes more and more narrow ($\rho_2 \rightarrow \rho_1$) the  energy levels tend to grow up (see Fig. \ref{fig:m=0-rho2} (right panel) in Appendix~\ref{sec:A}). This fact is easy to understand, bearing in mind that in the problem under consideration we have the interplay between the Landau and size quantization of the energy levels. When the disk becomes more and more narrow the first level of size quantization rises up like in the narrow quantum well with infinitely high walls.
As shown in Appendix~\ref{sec:B}, Eq.~(\ref{energy_levels}) for eigenenergies
acquires a much simpler form that involves a combination of the Bessel functions
of the first and second kind:
\begin{equation}
\label{energy_levels_Bessel}
\begin{split}
  {J_{\left| m \right|}}\left( {\sqrt {2\tilde \varepsilon } {\rho _1}} \right){Y_{\left| m \right|}}\left( {\sqrt {2\tilde \varepsilon } {\rho _2}} \right)  \\
   - {J_{\left| m \right|}}\left( {\sqrt {2\tilde \varepsilon } {\rho _2}} \right){Y_{\left| m \right|}}\left( {\sqrt {2\tilde \varepsilon } {\rho _1}} \right) = 0. 
\end{split}
\end{equation}
When the magnetic field disappears, $\hbar \omega_c \to 0$, then
$ \sqrt{2 \tilde \varepsilon} \rho_{1,2} \to r_{1,2} \sqrt{2 m_e E}/\hbar$, so that   
Eq.~(\ref{energy_levels_Bessel}) reduces to the well-known equation that describes, for example, motion of a free particle on the annulus \cite{Suzuki1996NC}.

The equation (\ref{energy_levels_Bessel}) can be solved in the high energy approximation (see Eq.~(\ref{narrow-spectrum}) in Appendix~\ref{sec:B}) what results in the following spectrum
\begin{equation}
\label{narrow-spectrum-units}
E_{n,m} = \frac{\hbar^2}{2 m_e}  \left( \frac{\pi^2 n^2}{d^2} + \frac{m^2 -1/4}{r^2}\right)
- \frac{\hbar \omega_c}{2} m
\end{equation}
with $n=1,2,\ldots,$ and $m = -\infty,\ldots, 0, \ldots, \infty$ and $d=r_2 - r_1 $ being the width of the annulus. 
We stress that both Eq.~(\ref{energy_levels_Bessel}) and Eq.~(\ref{narrow-spectrum-units})
are valid for $d \ll \sqrt{2 \pi} l \ll r$ (see Appendix~\ref{sec:B}).
The first term of Eq.~(\ref{narrow-spectrum-units}) is nothing else that the energy spectrum of the free  electron gas confined in quantum well of the width $d$. Writing Eq.~(\ref{narrow-spectrum-units})
we also neglected the term $\sim d^2$ which is present in
Eq.~(\ref{narrow-spectrum}) in Appendix \ref{sec:B} and replaced $r_2$ by $r=r_1 \approx r_2$.

We note that the spectrum (\ref{narrow-spectrum-units}) can be rewritten
in terms of the flux $\Phi = \pi r^2 B$ 
that goes through the ring
\begin{equation}
\label{narrow-spectrum-units1}
E_{n,m}\! =\! \frac{\hbar^2}{2 m_e} \left\{\frac{\pi^2 n^2}{d^2}\! +\! 
\frac{1}{r^2} 
\left[\left( m \!- \!\frac{\Phi}{\Phi_0}\right)^2
\!- \!\frac{\Phi^2}{\Phi_0^2}\! -\! \frac{1}{4}\right] \right\}.
\end{equation} 
When the disk becomes very narrow, $d \to0$, it is sufficient to restrict ourselves
by considering in Eqs.~(\ref{narrow-spectrum-units}) and (\ref{narrow-spectrum-units1}) only the lowest,
$n=1$ level. Then
one can see that the term $(m- \Phi/\Phi_0)^2$
coincides with the spectrum of the one-dimensional ring \cite{Gefen1}.
The extra terms $\Phi^2/\Phi_0^2 +1/4$ appear because of
the homogeneity of magnetic field in our model, while
the spectra in Refs.~\cite{Kulik1,Gefen1} were obtained for the ring threaded
by the flux. 

\subsubsection{Wave function}

The radial wave function corresponding to the
eigenenergies given by Eq.~(\ref{energy_levels_Bessel})
reads 
\begin{equation}
\label{radial-wf_Bessel}
\begin{split}
f_{nm}(\rho) = C \left[
  {{Y_{\left| m \right|}}\left( {\sqrt {2\tilde \varepsilon } {\rho _1}} \right)
  J_{\left| m \right|}}\left( {\sqrt {2\tilde \varepsilon } {\rho}} \right)  \right. \\
  \left. - {J_{\left| m \right|}}\left( {\sqrt {2\tilde \varepsilon } {\rho _1}} \right){Y_{\left| m \right|}}\left( {\sqrt {2\tilde \varepsilon } {\rho}} \right) \right]. 
\end{split}
\end{equation}

The last expression can be further simplified in the large energy limit supposing the validity of the conditions $r=r_1 \approx r_2 \gg l$. The latter inequality imposes the restriction on a magnetic field, which cannot be too weak:  $\omega_c \gg \hbar/(m_er^2)$. Physically this means that the quantization related to the magnetic field must dominate on the  size  quantization of the tangential electron motion along the ring. 
As result, the wave function acquires the form (\ref{asymptotic-wf}) which turns out to be more convenient for further calculation of the partial current  $I_{nm}$.

\subsubsection{Persistent currents in a large and narrow annulus}

Let us  now consider the current in a large and narrow annulus 
under the conditions $r=r_1 \approx r_2 \gg l$. 
The partial current $I_{nm}$ carried by  the state with definite quantum 
numbers $n,m$ can be written down 
by substituting the integral  (\ref{integral}) to the second line of Eq.~(\ref{I_nm})
\begin{equation}
\label{I_nm-narrow_disk}
\begin{split}
{I_m} \equiv  {I_{nm}} & = \frac{e \hbar}{2 \pi m_e} \left(\frac{m}{r^2} - \frac{1}{2 l^2}
\right)\\
& =\frac{e \hbar}{2 \pi m_e r^2} \left(m - \frac{\Phi}{\Phi_0} \right),
\end{split}
\end{equation}
where within our approximation for the wave function Eq.~(\ref{radial-wf_Bessel}) $I_{nm}$ turn out to be independent on the principal quantum number $n$, thereby ${I_{nm}} \equiv {I_m}$ (see details in Appendix \ref{sec:B}). 
The second line of Eq.~(\ref{I_nm-narrow_disk}) 
is identical to the corresponding expression
for the partial current in  \cite{Gefen1}, in spite of the fact that the given above 
spectrum (\ref{narrow-spectrum-units1}) contains the additional terms.

One can easily check that exactly the same expression for $I_{nm}$
follows directly from the Byers-Yang formula (\ref{Byers-Yang}) with 
the derived above spectrum (\ref{narrow-spectrum-units}).

The found spectrum (\ref{narrow-spectrum-units}) and wave function 
(\ref{radial-wf_Bessel}) in the above approximations 
will allow us to obtain the value of full current flowing in the narrow annulus vs its radius and width.

Substitution of Eq.~(\ref{radial-wf_Bessel}) into Eq.~(\ref{current_density_up1}) for the tangential component of the current density  and subsequent straightforward integration leads to the cumbersome formula that can be expressed in terms of sine and cosine integral functions. However, within our assumption about the large scale annulus ($r_1 \approx r_2 = \rho l$, $\rho \gg 1$) with the small width $d \lesssim l \sqrt{2\pi}$ ($\delta =d/l \lesssim \sqrt{2\pi}$) we can simplify the expression for the current to the form  [see 
Eqs.~(\ref{radial-integral}) - (\ref{integral}) in Appendix~\ref{sec:B}]
\begin{equation}
\label{current_narrow_disc_general_1}
\begin{split}
I  = \sum_{\substack{ n=1 } }^\infty  \sum_{\substack{m = -\infty } }^\infty I_{m} \theta(\mu-E_{n,m})
\end{split}
\end{equation}
The case of very narrow ring, when only one level of dimensional quantization occurs below the chemical potential, was studied by the authors of \cite{Gefen1,Kulik1}. We can reproduce their result accounting in Eq. (\ref{current_narrow_disc_general_1}) for only the lowest principal quantum number ($n=1$) and performing the remaining summation over azimuthal number by means of Poisson formula:
\begin{equation}
\label{Geffen_like}
I = \frac{{e\hbar k_F}}{{2\pi^2 {m_e}{r}}}\sum\limits_{k = 1}^\infty  {\sin } \left( {\frac{{2\pi k\Phi }}{{{\Phi _0}}}} \right)\frac{{\cos (2\pi k{k_F}r)}}{{ k}},
\end{equation}
where $k_F$  is the Fermi wave-vector.
Here is assumed that $\mu \gg \hbar \omega_c$ and $k_F r \gg 1$. The obtained magnitude of current oscillations can be compared to that one revealed in experiment \cite{Moler}. Taking $v_F = \SI{1.2e8}{cm/s}$
for gold, $r=\SI{670}{nm}$, one finds 
$I_0  \approx \SI{10}{nA}$  that is one order larger the observed values. 
Such overestimate is related to the fact that we do not consider accompanied effects such as temperature, disorder and the electron-electron interaction.

In Ref. \onlinecite{Richter} the problem of calculation of the persistent current in the large thin ring was considered in the semiclassical approximation basing on the Bohr-Sommerfeld quantization rule. This allowed the authors to avoid the sophisticated study of the Bessel function cross-product zeros (see Eq. (\ref{Bessel-cross-equation})) and get the explicit expression for the density of levels distribution in the simple form. With its use they obtained the complete analytical expression for the persistent current in the large ring accounting for the contributions of all semiclassical electron trajectories and, additionally, for the finite temperatures.

In case of the annulus of the finite width the averaging of Eq. (\ref{Geffen_like}) over the varying radius $r$ results in strong cancellation of oscillations, the total current strongly decreases. In order to get its value we can  perform both summations in Eq. (\ref{current_narrow_disc_general_1}) exactly. The presence of theta function in it implies the finite limits in both summations:
\begin{equation}
\label{current_narrow_disc_general_main}
\begin{split}
I =\frac{e \hbar}{4 \pi m_e l^2} \sum_{n=1}^{N_{max}} \sum_{m=M_{min}}^{M_{max}}  {\left( {\frac{2m}{\rho ^2}-1} \right)} .
\end{split}
\end{equation}
which are determined by the explicit expression for spectrum (\ref{narrow-spectrum-units}). 
The condition that the argument of theta-function in 
Eq.~(\ref{current_narrow_disc_general_main}) remains positive for fixed value of the chemical potential yields the constraints for the azimuthal quantum number $m$:
\begin{equation}
\begin{gathered}
  {M_{\min }} = \left[ { - \rho \sqrt {\frac{{2\mu }}{{\hbar {\omega _c}}} - \frac{{{\pi ^2}{n^2}}}{{{\delta ^2}}} + \frac{{{\rho ^2}}}{4}}  + {\rho ^2}} \right], \hfill \\
  {M_{\max }} = \left[ {\rho \sqrt {\frac{{2\mu }}{{\hbar {\omega _c}}} - \frac{{{\pi ^2}{n^2}}}{{{\delta ^2}}} + \frac{{{\rho ^2}}}{4}}  + {\rho ^2}} \right]. \hfill \\ 
\end{gathered} 
\label{Mmin_max}
\end{equation}
with $[\ldots]$ denoting the integer part.

What concerns the summation over the principal quantum number its upper limit $N_{max}$  can be determined from the condition of the positiveness of the square root in Eq.~({\ref{Mmin_max}}):
\begin{equation}
{N_{\max }} = \left[ {\frac{\delta }{\pi }\sqrt {\frac{{2\mu }}{{\hbar {\omega _c}}} + \frac{{{\rho ^2}}}{4}} } \right].
\label{Nmax}
\end{equation}
In the following we assume that   $N_{max} \gg 1$, i.e. the width of the disk is not too small and is limited by the conditions  $ {\pi {{\left( {2\mu /(\hbar {\omega _c}) + {\rho ^2}/4} \right)}^{ - 1/2}}} \ll \delta \lesssim \sqrt{2 \pi}$. If this were not so, $ N_ {max} $ would become less than $ 1 $ and because of such a strong size quantization  there would be no level left under the chemical potential. 

The summation over  $m$ in 
Eq.~(\ref{current_narrow_disc_general_main}) is trivial and results in
\begin{equation}
I = \frac{{e\hbar }}{{4\pi {m_e}{l^2}}}\sum\limits_{n = 1}^{{N_{max}}} {\left[ {\frac{{2\pi \rho }}{\delta }\sqrt {N_{\max }^2 - {n^2}}  + 1} \right]}.
\label{Igenring}
\end{equation}

\begin{figure}
\includegraphics[width=0.99\columnwidth]{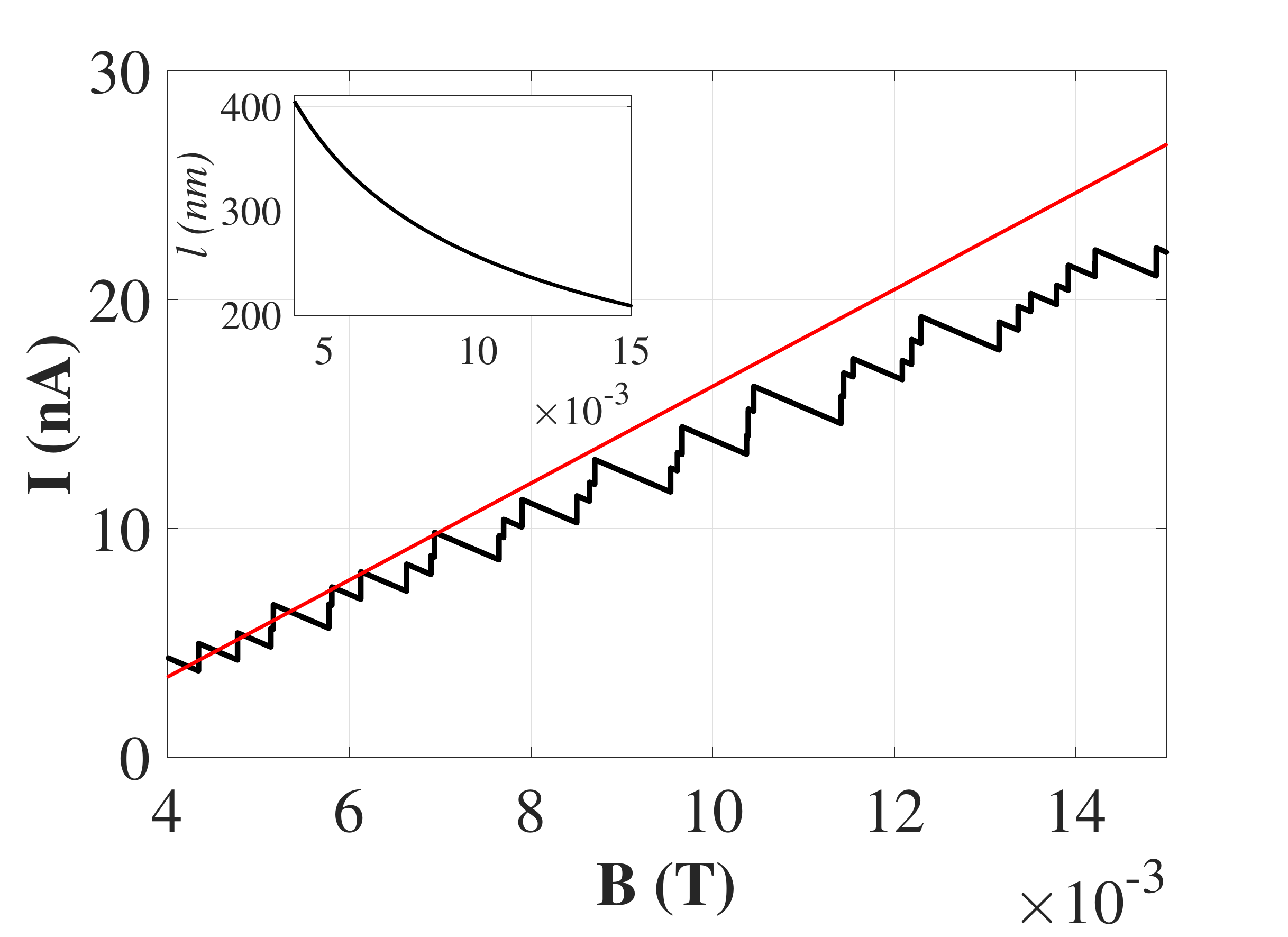}
\caption{Total current flowing in the narrow annulus ($d=50\, nm$) versus the strength of magnetic field. The radius $r=600\,nm$, concentration of the 2DEG is $10^{12}cm^{-2}$. The red line illustrates a linear dependence on the strength of a weak magnetic field in accordance with Eq. (\ref{current_narrow_disc_large_mu}).  The magnetic length remains larger than the annulus width (see inset).}
\label{IvsB}
\end{figure}

The characteristic dependence of the total current flowing in the narrow ring versus the strength of magnetic field is shown in Fig. \ref{IvsB}. One can see, that due to the presence of the integer part in Eq. (\ref{Nmax}), this dependence keeps saw-like character imposed on the general  growth.

To analyse the various regimes and in view of $N_{max} \gg 1 $ the summation over $n$ in Eq. (\ref{Igenring}) can be performed  applying the Euler–Maclaurin formula, what leads to:
\begin{equation}
\label{current_narrow_disc_final}
\begin{split}
 I = \frac{{e\hbar \rho}}{{4 {m_e}{ \delta l^2}  }} & \left[ {\frac{{\pi N_{\max }^2}}{2} - N_{\max }^2\arcsin \left( {\frac{1}{{{N_{\max }}}}} \right)  } \right.  \\
 &\left. - \sqrt {N_{\max }^2 - 1} + \frac{\delta}{\pi \rho}N_{max} \right].  
\end{split}
\end{equation}
Keeping the leading term of Eq.~(\ref{current_narrow_disc_final}) and substituting in it  Eq.~(\ref{Nmax}) we obtain the explicit expression for the current enveloping curve:
\begin{equation}
\label{current_narrow_disc_final2}
I = \frac{{e\hbar rd}}{{8\pi {m_e}{l^4}}}\left( {\frac{{2\mu }}{{\hbar {\omega _c}}} + \frac{{{r^2}}}{{4{l^2}}}} \right).
\end{equation}

We can analyze two different regimes of Eq. (\ref{current_narrow_disc_final2}).

\begin{itemize}

\item $ \hbar/(m_er^2) \ll \omega_c \ll  \sqrt{\mu  /(m_e r^2)} $. In this limit of weak enough fields,  the magnitude of the current increases linearly with growth of magnetic field:
\begin{equation}
\label{current_narrow_disc_large_mu}
I = \frac{e^2rd\mu}{4\pi c \hbar^2 }B=\frac{m_ee^2rd}{4c}n_{2D}B.
\end{equation}

\item  $ \sqrt{\mu  /(m_e r^2)} \ll \omega_c \ll \mu/\hbar $. In this interval the magnetic field becomes strong enough, though still remaining below the ultra-quantum limit (when the second term under the square root in Eq.~(\ref{Nmax}) starts to dominate on the first one); Eq.~(\ref{current_narrow_disc_final2}) reduces to
\begin{equation}
\label{current_narrow_disc_large_r}
I = \frac{{e^4{r^3}}d}{{32\pi c^3\hbar^2 {m_e}}}B^3,
\end{equation}
 and the magnitude of the persistent current increases more rapidly ($I \sim B^3$ instead of $I \sim B$).
\end {itemize}
 
\begin{figure}
\includegraphics[width=0.99\columnwidth]{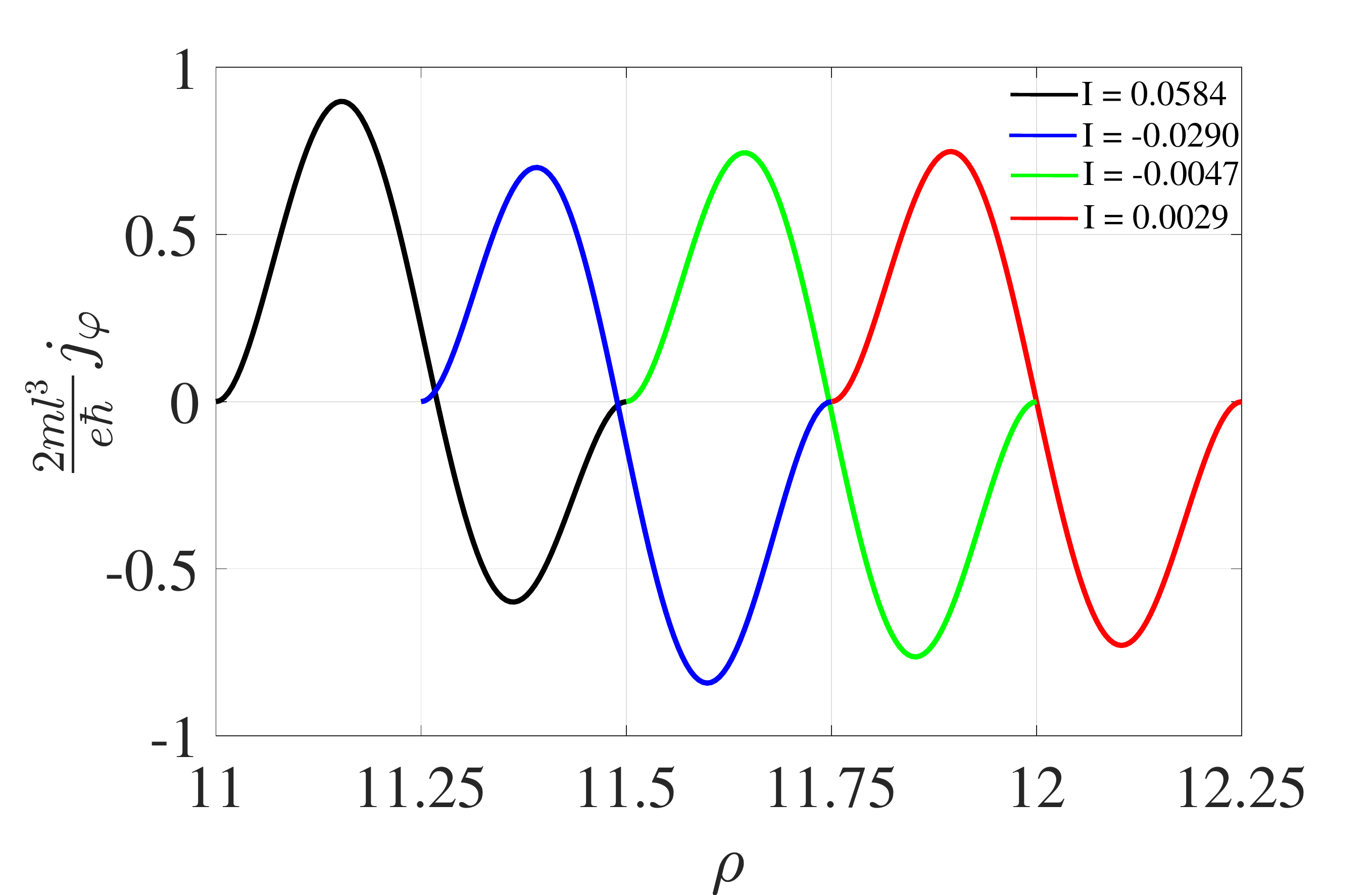}
\caption{Current density as a function of a dimensionless radial coordinate for a narrow annulus with a fixed width $d = 0.5$ and different inner and outer radii  $\rho_1 =11$, $\rho_2 =11.5$ (black) $\rho_1 =11.25$, $\rho_2 =11.75$ (blue), $\rho_1 =11.5$, $\rho_2 =12$ (green) and  $\rho_1 =11.75$, $\rho_2 =12.25$ (red). The chemical potential is chosen to be $\mu=20 \hbar \omega_c$ and $T=0$. Inset shows the values of the current in units of $\frac{{2{m_e}{l^2}}}{{\pi e\hbar }}$ in an annulus, where the color lines correspond to an appropriate current density profile of the system.}
\label{current_density_large_mu}
\end{figure}

Now we proceed to the discussion of the current radial distribution over the narrow annnulus under the same conditions $d  \lesssim \sqrt{2 \pi}l \ll r$ and $\hbar \omega_c \ll \mu$ ($N_{max} \gg1$). It can be found by means of the numerical analysis of Eq.~(\ref{current_density_up1}) with the wave functions and spectrum determined by Eqs.~(\ref{solution_radial}) and (\ref{energy_levels}). As one can see from Fig.~\ref{current_density_large_mu} the current density changes its sign as one moves from the inner to the outer edge. Moreover, one can notice that the total current $I$ (the integral of the current density) also changes its value and direction with the growth of the annulus radii (see the values in the upper right corner of Fig.~\ref{current_density_large_mu}). The latter fact is in accordance with the observed alterations of the current values in Fig.~\ref{IvsB} as a function of the magnetic field.  

\subsection{The spectral problem and persistent current in a small annulus with the infinitesimal inner radius} 

\subsubsection{Energy spectrum and wave function}

In the case when the inner radius of an annulus is smaller than the magnetic length one can forget about it and approximate the disk by the  solid one without hole in the centre at all (${\rho _1}=0$).  The  energy spectrum  in such a case is determined by the zeroes of the confluent hypergeometric function (of the first kind or Kummer's confluent hypergeometric function) (see e.g. Ref.~\cite{Rensink}):
\begin{equation}
\label{energy_levels_small_disc}
\Phi \left( {\frac{1}{2} - \varepsilon_{nm}  + \frac{{\left| m \right| - m}}{2},\left| m \right| + 1, \frac{\rho _2^2}{2}} \right) = 0.
\end{equation}

Being interested in the properties of a ``small'' disk we assume that its the outer  radius is smaller than the  magnetic length: 
$r_2 \lesssim l$ (weak magnetic field approximation). 
Corresponding spectrum acquires the form
(cf.  Eq.~(\ref{narrow-spectrum-units}))
\begin{equation}
\label{energy_levels_small_disc_new}
{E_{n,m}} = \frac{\hbar^2}{2 m_e} \frac{{j_{nm}^2}}{{r_2^2}} - 
\frac{\hbar \omega_c}{2} m,
\end{equation}
where ${j_{nm}}$ is the n-th zero of ${J_{\left| m \right|}}\left( z \right)$.

The radial component of the wave function can be also evaluated  
(see Eqs.~(\ref{hypergeom_Bessel_expansion})-(\ref{C_small_Corbino_disc}) in Appendix \ref{sec:C})
\begin{equation}
\label{radial_wave_func_small_disc}
f_{nm}\left( r \right) = \frac{1} {\sqrt{\pi}
{r_2 {J_{\left| m \right| + 1}}\left( {{j_{nm}}} \right)}}{J_{\left| m \right|}}\left( {{j_{nm}}\frac{r }{{{r_2}}}} \right).
\end{equation}
As in the case of the previous subsection these analytical findings will allow us to study the nontrivial full current genesis vs the size of such a small annulus.

\subsubsection{Persistent currents}

\begin{figure}
\includegraphics[width=0.99\columnwidth]{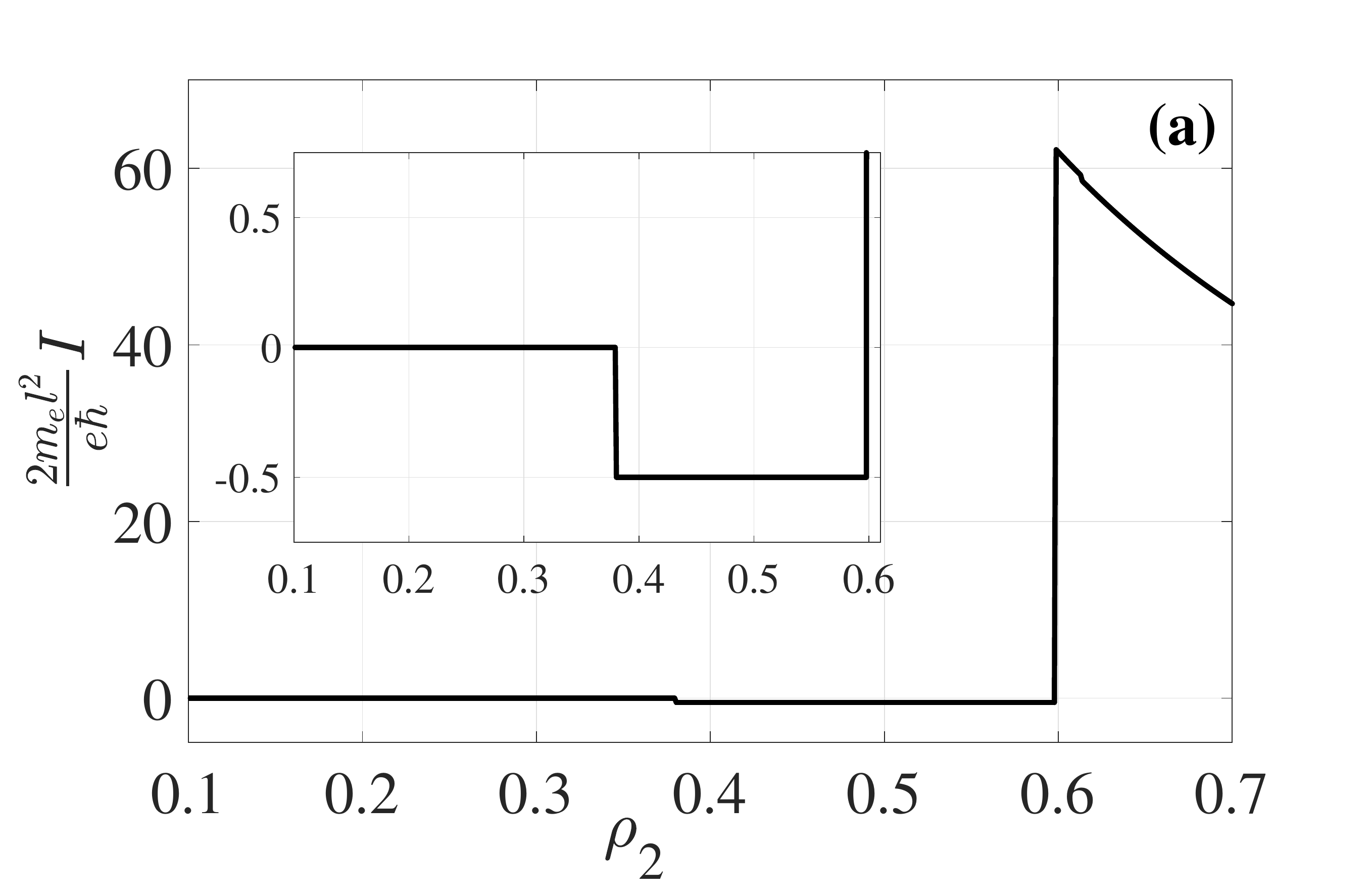}
\includegraphics[width=0.49\columnwidth]{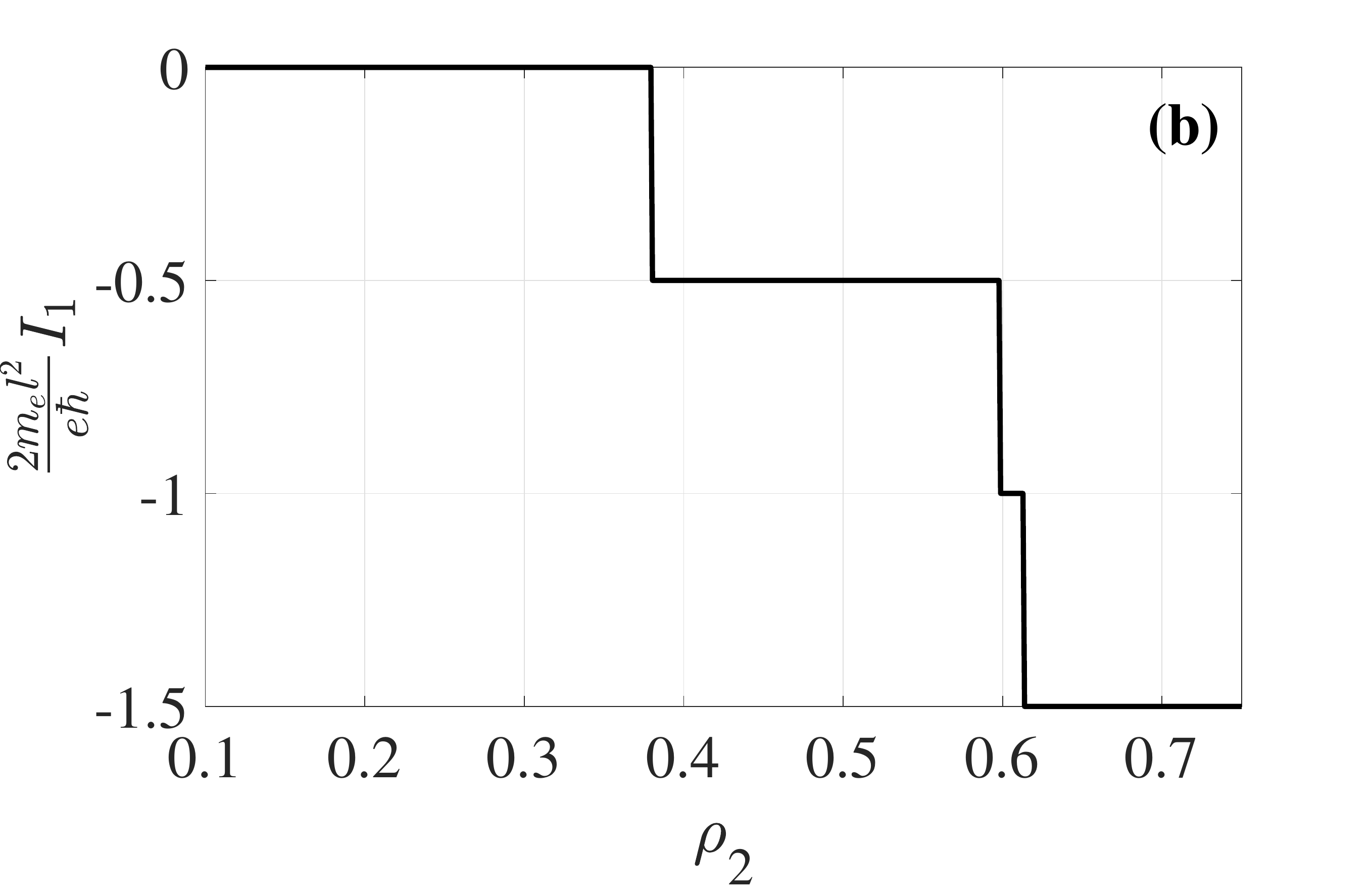}
\includegraphics[width=0.49\columnwidth]{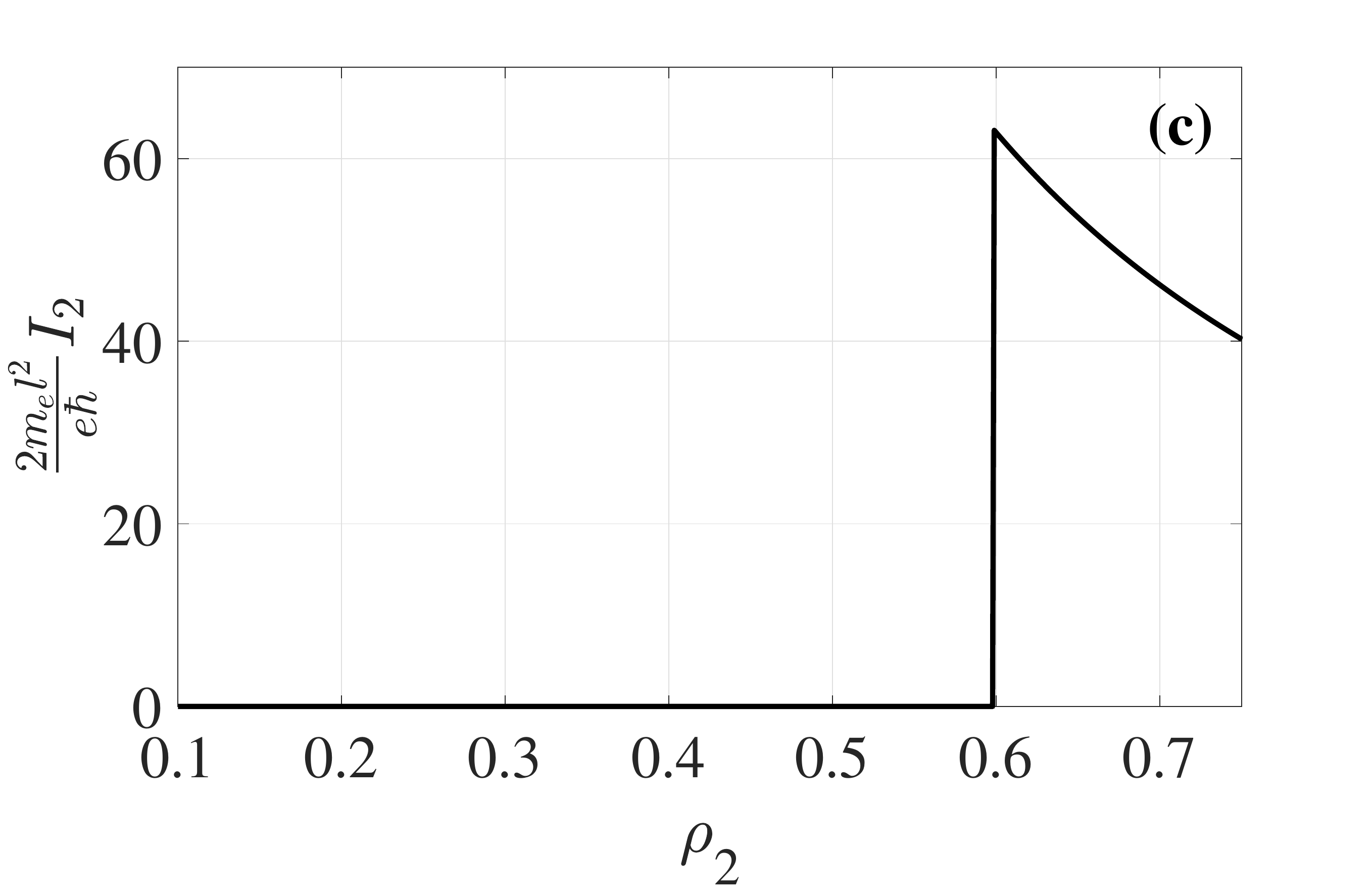}
\caption{(a) The full current as a function of a dimensionless radius $\rho_2$ for a small annulus with a infinitesimal $\rho_1$. The chemical potential is equal to $\mu=20 \hbar \omega_c$ and $T=0$. Inset in (a) shows the zoom of the plot, where the current has negative values. (b) The negative contribution $I_1$ to the full current, while (c) is the positive contribution $I_2$.}
\label{current_small_disc_figure}
\end{figure}

In the case of a  disk of the outer radius $r_2 \lesssim l$ 
with the hole much smaller magnetic length
($r_1 \to 0$) the total current can be determined by 
Eq.~(\ref{current_density_up1}) for the tangential component of the current density and corresponding expression for radial part of the wave-function (\ref{radial_wave_func_small_disc}).
Integration can be explicitly performed in terms of the
Bessel functions (see Appendix~\ref{sec:C}):
\begin{equation}
\label{current_small_disc_main}
I = I_1 + I_2.
\end{equation}
Here 
\begin{equation}
\label{current_small_disc_main_1}
{I_1} =  - \frac{{e\hbar }}{{4 \pi {m_e}{l^2}}}\sum_{\substack{m = -\infty\\ n=1 } }^\infty  \theta(\mu- E_{n,m})
\end{equation}
is the diamagnetic current
and 
\begin{equation}
\label{current_small_disc_main_2}
I_2 = \frac{e \hbar}{2 \pi m_e r_2^2} \sum_{\substack{m = -\infty\\ n=1 } }^\infty 
\frac{A_{nm} \mbox{sgn} (m) }{J^2_{|m|+1} (j_{nm})} \theta(\mu- E_{n,m})
\end{equation}
is the paramagetic current with
$A_{nm}$ given by Eq.~(\ref{A_nm})
and $\mbox{sgn} (0)=0$.

\subsubsection{Derivation using Byers-Yang formula}

One can also verify 
that Eqs.~(\ref{current_small_disc_main}) -
(\ref{current_small_disc_main_2})  follow directly from
the Byers-Yang formula (\ref{Byers-Yang}). 
Differentiating  the spectrum (\ref{energy_levels_small_disc_new})  
one obtains
\begin{equation}
\label{Inm-derivative-bessel}
I_{nm}=\frac{e\hbar}{2m_e}\left(\frac{{\rm sign}(m)}{\pi r_2^2}j_{nm}
\frac{\partial j_{n\nu}}{\partial\nu}\Big|_{\nu=|m|}-\frac{1}{2\pi l^2}\right),
\end{equation}
where it was taken into account that roots $j_{nm}$ of the equation $J_{|m|}(z)=0$ 
depend on $|m|$. Substituting the derivative of these roots $\partial j_{nm}/ \partial|m|$
[see  Eq.~(\ref{root-bessel-derivative}) in Appendix~\ref{sec:C}]
in Eq.~(\ref{Inm-derivative-bessel})
 we arrive at the final result
\begin{equation}
\label{I_nm-disk}
I_{nm}=\frac{e\hbar}{2m_e}\left(\frac{A_{nm}{\rm sign}(m)}{\pi r_2^2J^2_{|m|+1}(j_{nm})}-\frac{1}{2\pi l^2}\right).
\end{equation}
As one can easily see, the second term of Eq.~(\ref{I_nm-disk}) corresponds
to $I_1$ and the first term to $I_2$, respectively.

\subsubsection{Emergence of the current states in the small annulus}

The numerical simulation of Eq.~(\ref{current_small_disc_main}) is represented in Figure \ref{current_small_disc_figure}. With the chemical potential equal to $20\hbar \omega_c$ and in the vicinity of the zero temperature two leaps of the current are clearly observed. The behaviour of these leaps can be easily understood if we consider separately two contributions 
of the currents $I_1$ and $I_2$, where $I_1$ given by Eq.~(\ref{current_small_disc_main_1}) is the negative and independent of the radius $\rho_2$ and $I_2$ represented by Eq.~(\ref{current_small_disc_main_2}) is inversely proportional to the value of outer radius.
As long as all energy levels Eq.~(\ref{energy_levels_small_disc_new}) of the small annulus are located above the given value of the chemical potential there is no current in an annulus (see inset in Fig.~\ref{current_small_disc_figure}a). The first leap is connected with the emergence of the quantum state with $m=0$ below the chemical potential. Due to this the full current dependence has only one contribution from $I_1$ given by Eq.~(\ref{current_small_disc_main_1}) with zero part from $I_2$ and, therefore, is the constant until the certain value of $\rho_2$ 
(Fig.~\ref{current_small_disc_figure}~b), where another energy levels with the nonzero azimuthal number
$m$ give rise a new leap and a new constant. 

Further increase of the radius allows to involve other energy levels with $m \ne 0$. This leads to the activation of the second contribution $I_2$ given by Eq.~(\ref{current_small_disc_main_2}).
As a result together with the second leap in $I_1$ (Fig.~\ref{current_small_disc_figure}b) similar effect occurs for $I_2$ 
(Fig.~\ref{current_small_disc_figure}c).  

\section{Annulus of the arbitrary sizes: numerical analysis}

In principle, the expression for the current density (\ref{current_density_up1}) allows to investigate its profile for the disk of arbitrary  radii (in practice the computation for a wide enough disk is very time-consuming).

\begin{figure}
\includegraphics[width=0.99\columnwidth]{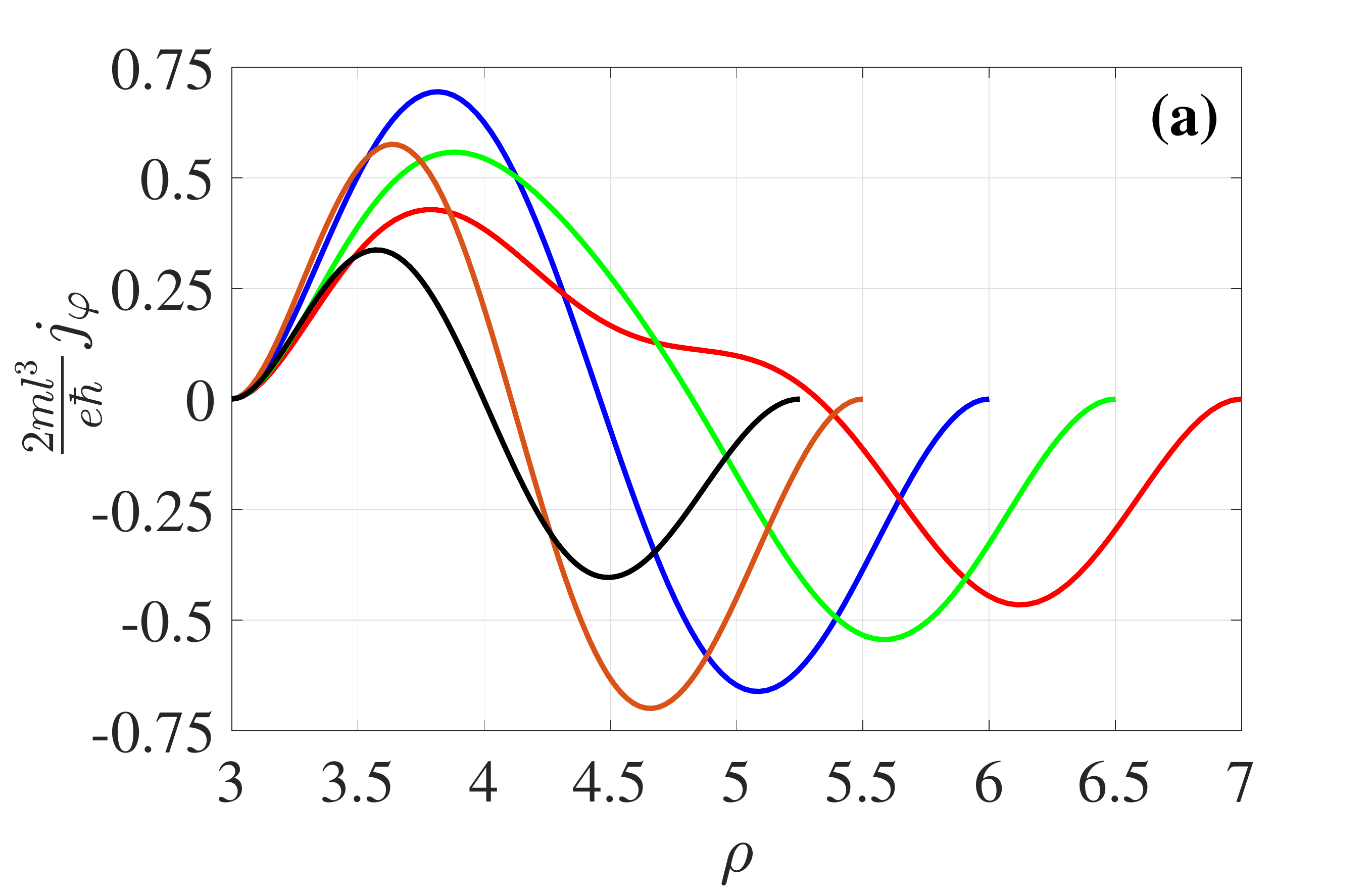}
\includegraphics[width=0.99\columnwidth]{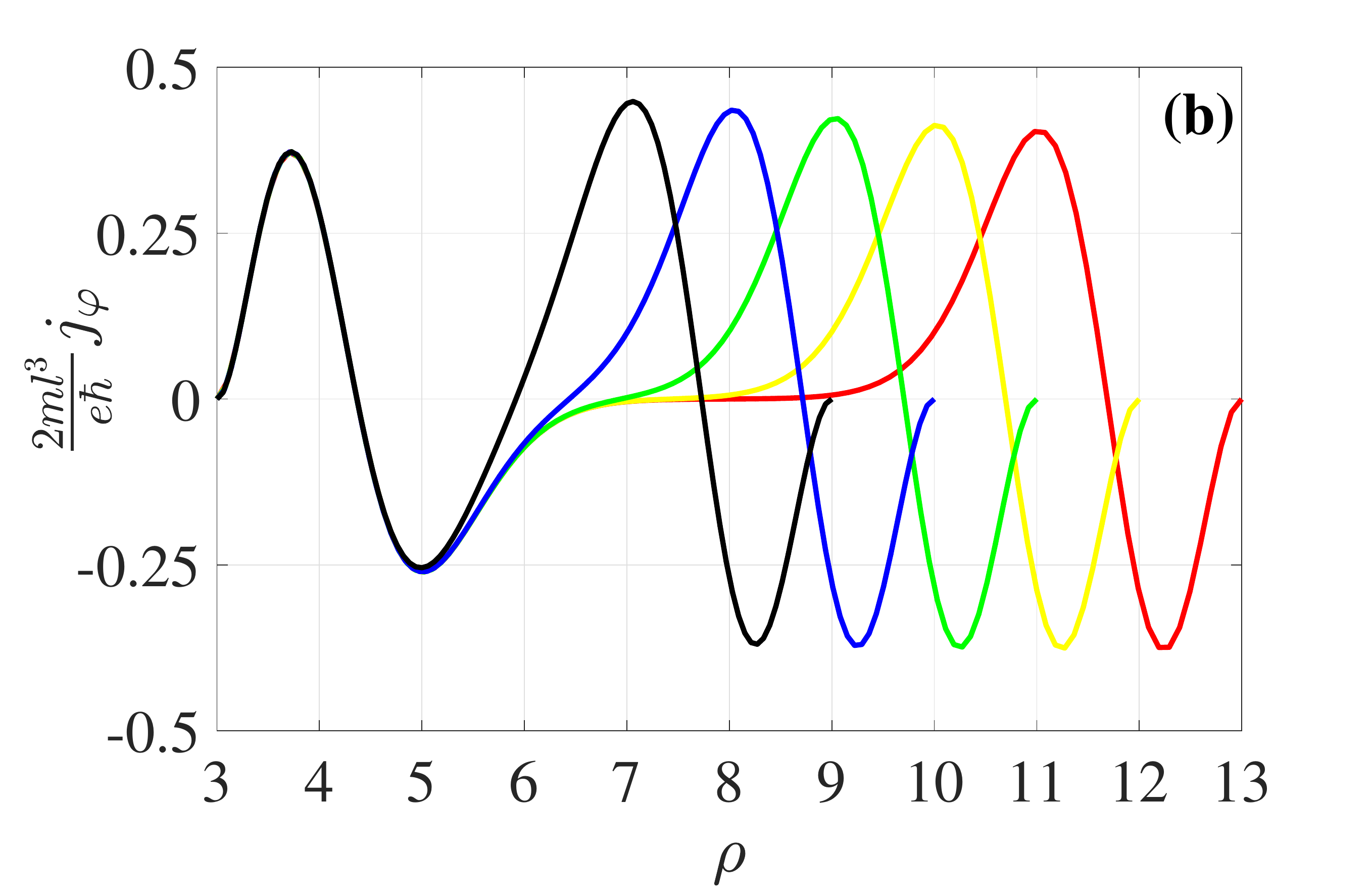}
\caption{Current density as a function of a dimensionless radial coordinate  for an annulus with a fixed inner radius $\rho_1 = 3$ and different outer radii  $\rho_2 =5.25$ (black) $\rho_2 =5.5$ (brown), $\rho_2 =6$ (blue), $\rho_2 =6.5$ (green) and  $\rho_2 =7$ (red) in (a) and $\rho_2 =9$ (black), $\rho_2 =10$ (blue), $\rho_2 =11$ (green), $\rho_2 =12$ (yellow) and  $\rho_2 =13$ (red) in (b). The chemical potential is chosen to be $\mu=1.1 \hbar \omega_c$ and $T=0$.}
\label{current_density_profiles}
\end{figure}

\begin{figure}
\includegraphics[width=0.99\columnwidth]{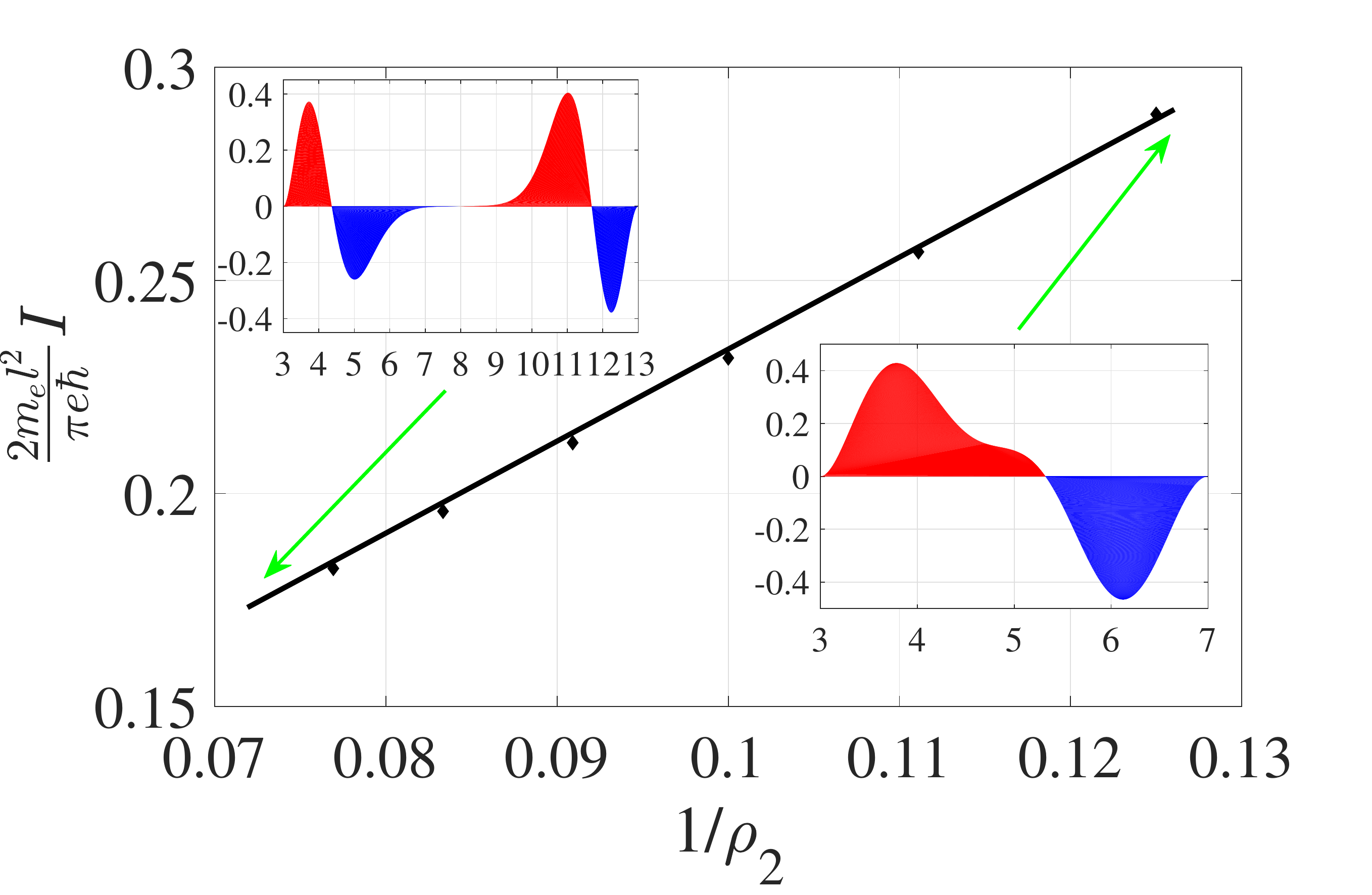}
\caption{Total current as a function of a dimensionless inverse radial coordinate $1/\rho_2$ for an annulus with a fixed inner radius $\rho_1 = 3$. The chemical potential is equal to $\mu=1.1 \hbar \omega_c$ and $T=0$. Insets show current density profiles for $\rho_2=7$ and $\rho_2=13$.}
\label{full_current_figure}
\end{figure}

We studied the current density as a function of a radial coordinate  for an annulus with a fixed inner radius $\rho_1 = 3$ and the set of outer radii from $\rho_2 =5.25$ to $\rho_2 =13.0$. The corresponding results are presented in Figures \ref{current_density_profiles} and \ref{full_current_figure}.

Figure~\ref{current_density_profiles} shows the the current density as the function of the dimensionless radial coordinate of an annulus $\rho$ for  the fixed inner radius $\rho_1=3$. Relatively narrow annulus shows almost 
a sinusoidal type of the behaviour with the maximal positive value of the current  density near the inner radius and the minimal negative one near the outer edge of a system (Fig.~\ref{current_density_profiles}a, black line). 

With the increasing of the width disk (the same with the increasing of the outer radius) 
the deformation of the current density profile is begun. One can see in Figure \ref{current_density_profiles}~a (red line) that in the middle part of the annulus the current density profile starts to be flattened with the zero value. 
This suppression is clearly seen in Fig.~\ref{current_density_profiles}, where the evolution of the current density profile for the wide disk is shown. Such a behaviour can be easily understood from the energy spectrum for a wide disk when electrons strive to approach Landau levels as it can be seen in Fig. \ref{enery_level_general}~b and as a result make infinitesimal contributions to the current density distribution inside the disk. Moreover, together with the flattened zero part of $\j_{\varphi}$ 
the double changing of the current density sign is observed near in the vicinity of the inner and outer edge of the annulus.  

The occurrence of such complicated current density profiles should be taken into account in experiments with persistent currents in a ring, where as we have shown already in Fig. \ref{current_density_profiles} the local magnetic response of a system can be changed significantly even for a ring with the relatively small width.

The numerical calculations of  the full current as the
function of the inverse $\rho_2$ are presented in Fig.~\ref{full_current_figure}. 
They were obtained by numerical integration of the previously obtained current 
densities for  the fixed radius $\rho_1$.
The dependence of total current on $\rho_2^{-1}$ 
exhibits unambiguously the decay of the persistent current with the increase of the outer radius of the disk. 
This result is not surprising and is in agreement with Eq. (\ref{Inmlast_update}), obtained from Byers-Yang formula.

\section{Conclusions}

We have presented a comprehensive analytic and numeric study of the electron spectra 
and occurrence of persistent current in the 2DEG  filling the annulus of arbitrary dimensions subjected to constant magnetic field. The results obtained in this article can be summarized as follows.

i)  {\bf The case of a nanodot}. When the outer radius of the annulus is small
with respect to the magnetic length, $r_2 \lesssim l$, while that one of the inner hole is infinitesimaly small  (see Fig.~{\ref{sketch_energy_sol}d})  no current flows in the system. The only available states here correspond to zero azimuthal number which do not carry paramagnetic current (see  
Eq.~(\ref{current_small_disc_main_2}) and take into account that $A_{n0} =0$).
The diamagnetic contribution is also absent, because $E_{n,0} > \mu$.
As the outer radius increases, we observe how the first current state appears simultaneously with the emergence of the first  $m \neq 0$ state 
(see Fig.~\ref{current_small_disc_figure}). 

ii)  {\bf The case of a narrow annulus.} This is another case amendable for the analytic solution: i.e. when the annulus radius $r$ is much larger than the magnetic length, while its width is less or comparable with it ($d \lesssim \sqrt{2 \pi}l \ll r $, see Fig. {\ref{sketch_energy_sol}}c) and the magnetic field is not ultraquantum ($\hbar \omega_c \ll \mu$). 

When the ring is very narrow and only one level of radial dimensional quantization occurs below the chemical potential (i.e. $d \lesssim k_F^{-1} =\hbar/\sqrt{2m_e \mu}$), we reproduce the persistent current oscillations occurring in a nanoring \cite{Gefen1,Kulik1}.  Yet, our general expression for current Eq. (\ref{current_narrow_disc_general_1}) together with the spectrum (\ref{narrow-spectrum-units1}) allows also to analyze the multi-channel case of a relatively wide annulus. The obtained Eq. (\ref{Igenring}) reproduces the vanishing saw-like oscillations at the background of the persistent current growing with the increase of magnetic field (see Fig. \ref{IvsB} and Eqs. (\ref{current_narrow_disc_large_mu}) and (\ref{current_narrow_disc_large_r})). 

The analysis of Eq.~(\ref{current_density_up1}) allows us to study the radial distribution of the current density over the narrow annnulus. As one can see from Fig.~\ref{current_density_large_mu} the current density changes its sign as one moves from the inner to the outer edge. Moreover, the total current $I$ also changes its value and direction with the growth of the annulus radii. The numerical analysis presented in Fig.~\ref{current_density_large_mu} allows to see the alteration of the current direction frequently observed in experiment 
with nanorings (see Ref.~\cite{Moler}).

 iii) {\bf The case of a wide disc.}  We have succeeded to make a considerable progress in study of the general case of the annulus of arbitrary dimensions  $r_2 > r_1 \gtrsim l$ (wide disc with a hole in the center). Detailed analysis of the spectral problem for the edge states resulted in 
Eqs.~(\ref{energy_levels_k=inf_r1}),  (\ref{energy_levels_k=inf_r2}) 
(Fig.~{\ref{sketch_energy_sol}b}).
Further application of the Byers-Yang formula allowed us to reveal that the 
total current in a wide disk is inversely proportional to the strength of magnetic field and 
is determined by the difference in curvatures ($r_1^{-1}-r_2^{-1}$) of the 
inner and outer edges [see Eq.~(\ref{Inmlast_update})]. This clearly demonstrates 
its fading away in the case of the standard rectangular geometry 
(see Refs.~\cite{Heuser,MacDonald1984PRB}) and the annular geometry but
neglecting the curvature effects \cite{Halperin1982PRB,Varlamov}. Moreover, 
we have revealed the structure of current density profiles in such a geometry. This information can be important for the understanding of future non-invasive experiments (by means of scanning SQUID microscope) studied the magnetic response in a wide disc.

The obtained results allowed us to apply them for the analysis of some experimental findings (see Ref.~\cite{Moler}) and to establish very reasonable coincidence 
[see the estimates after Eq.~(\ref{Inmlast_update})]. Yet,  their validity can be restricted 
both by disorder and by electron-electron interaction (see e.g.~\cite{Glazman1992}.)  The latter becomes
noticeable for the electronic states at the almost empty quantized levels rounding for example the abrupt teeth in Fig. \ref{IvsB}.

Recently a large progress was achieved in the high-resolution (on the submicrometer scale) imaging
of the magnetic field and reconstructing current density distribution in graphene ribbons
\cite{Tetienne2017ScAdv}. Furthermore there is a hope that the same approach can be applied
to thin film systems. We stress that the considered 
homogeneous magnetic field geometry corresponds to the real experimental conditions better 
than the Aharonov-Bohm flux penetrating the ring.

Our study points out clear evidence of the geometry significance for more precise and accurate interpretations of experiments with persistent current density distribution in  mesoscopic rings and similar systems.

\begin{acknowledgments}
V.P.G. and S.G.Sh. acknowledge a support by the National Research Foundation of Ukraine grant  (2020.02/0051) "Topological phases of matter and excitations in Dirac materials, Josephson junctions and magnets". Y.Y. acknowledges support by the CarESS project. A.A.V. is grateful to Yu.~Galperin, A.~Kavokin, and V.B.~Shikin for valuable discussions. S.G.Sh. thanks V.~Kagalovsky for useful discussion. The authors  are grateful to C.~Petrillo for critical reading of the manuscript and valuable comments.

\end{acknowledgments}	

\begin{widetext}
\appendix
\section{Derivation of asymptotic expressions for the eigenvalue problem}
\label{sec:A}

\subsection{Alternative form of the equation for eigenenergies of the Corbino disk}

Using the relation (in notations of \cite{Bateman1}) between Whittaker functions and confluent hypergeometric functions of the first $\Phi(a,b,z)$ and the second kind $\Psi(a,b,z)$, respectively, Eq.~(\ref{energy_levels}) can be written the following form
\ba
\label{eq:confluent-functions}
&&\Psi\left(\frac{1}{2}-\epsilon+\frac{|m|-m}{2},|k|+1;\frac{\rho_1^2}{2}\right)\Phi\left(\frac{1}{2}-\epsilon+\frac{|m|-m}{2},|m|+1;\frac{\rho_2^2}{2}\right)
\nonumber\\
&&-\Phi\left(\frac{1}{2}-\epsilon+\frac{|m|-m}{2},|m|+1;\frac{\rho_1^2}{2}\right)\Psi\left(\frac{1}{2}-\epsilon+\frac{|m|-m}{2},|m|+1;\frac{\rho_2^2}{2}\right)
=0.
\ea
This form turns out to be useful for finding different analytical asymptotic solutions in some limits. 

\subsection{Eigenvalue equation for the disk with large outer radius}

We start investigation of the energy levels distribution  from the case of the disk with large outer radius $\rho_2 \gg 1$ and  
arbitrary inner one $\rho_1 < \rho_2 $.
The general Eq.~(\ref{energy_levels}) in this case can be simplified using the asymptotic expressions 
for Whittaker functions for large arguments \cite{Abramowitz},
\begin{equation}
\label{assymptot_M}
{M_{\tilde\varepsilon ,\frac{{\left| m \right|}}{2}}}\left( {\frac{\rho^2}{2}} \right) \approx 
{e^{\rho^2/4}}{\left( {\frac{\rho ^2}{2}} \right)^{ - \tilde\varepsilon }} 
\left[
\frac{{\Gamma \left( {1 + \left| m \right|} \right)}}{{\Gamma \left( {\frac{1}{2} + \frac{1}{2}\left| m \right| - \tilde\varepsilon } \right)}} + O \left( \frac{1}{\rho^2}\right) \right],
\end{equation}
and
\begin{equation}
\label{assymptot_W}
{W_{\tilde\varepsilon ,\frac{{\left| m \right|}}{2}}}\left( {\frac{\rho ^2}{2}} \right) \approx 
{e^{-\rho^2/4}}{\left( {\frac{\rho ^2}{2}} \right)^{  \tilde\varepsilon }} 
\left[1 + O \left( \frac{1}{\rho^2}\right) \right].
\end{equation}
In result one arrive to the following equation for the energy levels
\begin{equation}
\label{energy_levels_thin_new}
{W_{\tilde\varepsilon ,\frac{{\left| m \right|}}{2}}}\left( {\frac{\rho_1^2}{2}} \right) = 0.
\end{equation}
In other words, the energy dispersion relation for a Corbino disk with the very large external radius and fixed internal radius is determined by zeros of the Whittaker function. It is worth to mention that the roots of Eq.~(\ref{energy_levels_thin_new}) can be approximated by those ones of Bessel or Airy functions (see, e.g. \cite{Gabutti}).
The numerical solutions of Eq.~(\ref{energy_levels_thin_new}) are presented in Fig.~\ref{fig:infinite-rho-2}.
\begin{figure}
\includegraphics[width=0.49\columnwidth]{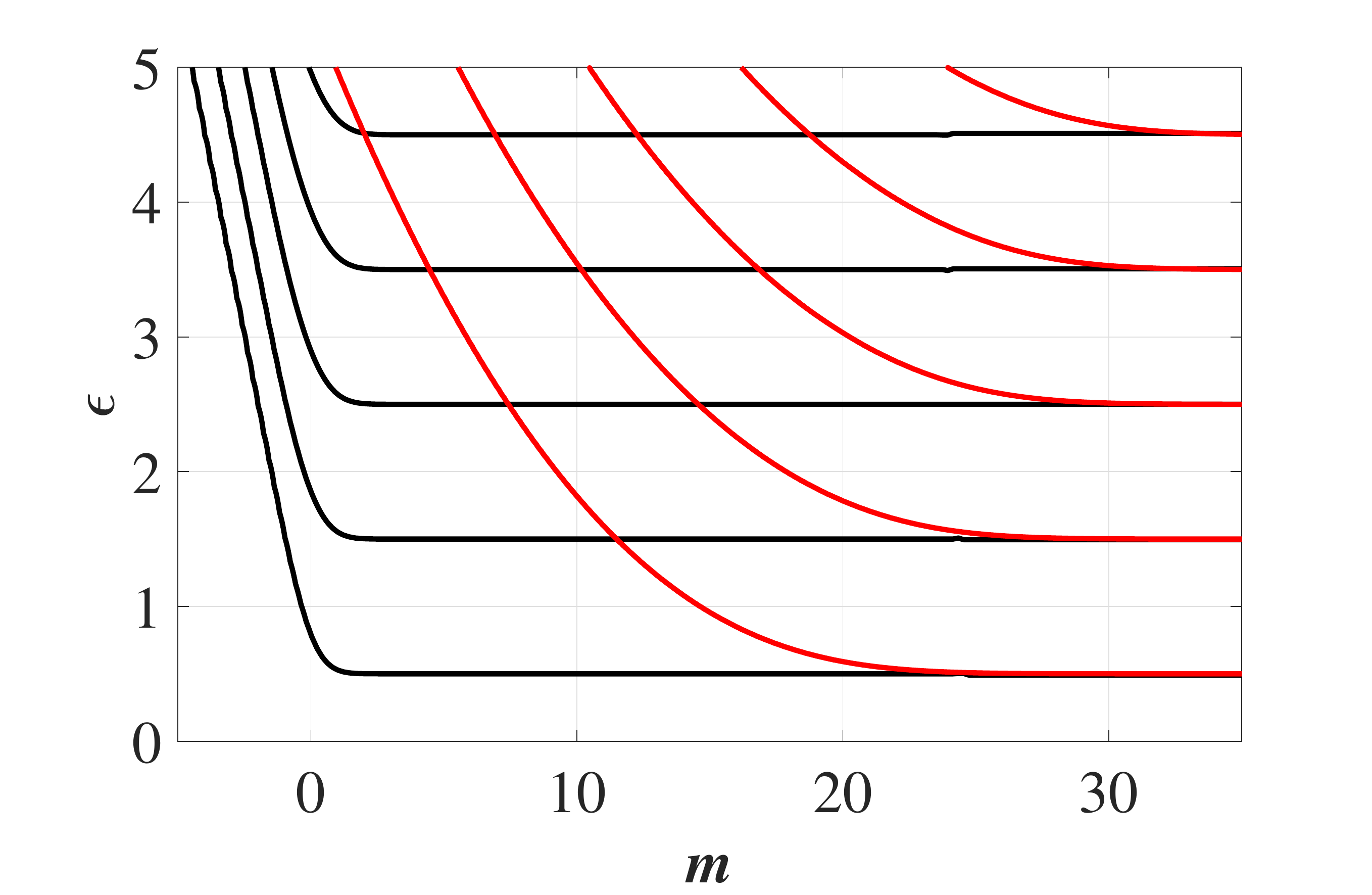}
\includegraphics[width=0.49\columnwidth]{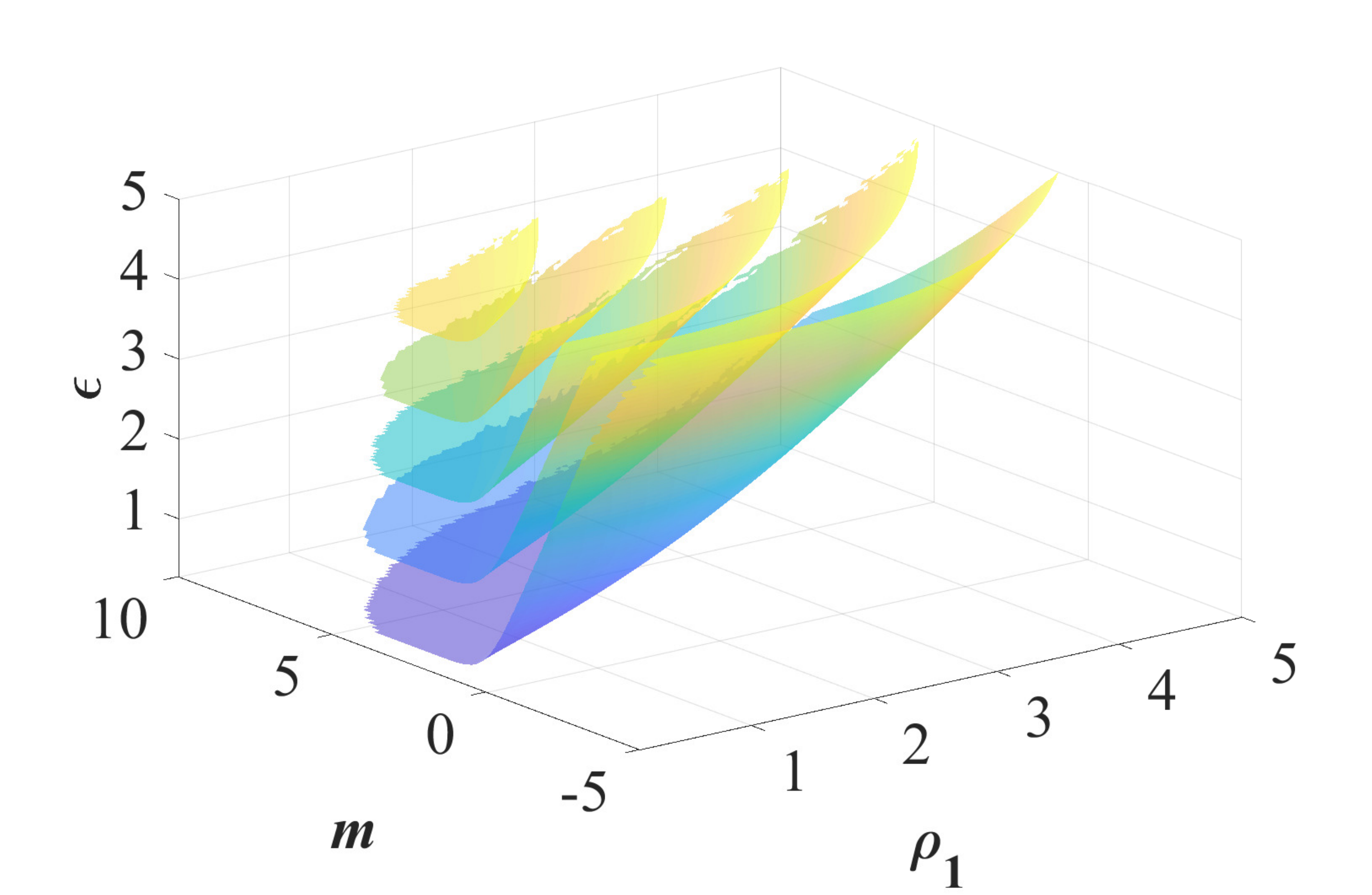}
\caption{(Left) The energy levels $\epsilon_{n,m}$ of the very large disk (see Eq.~(\ref{energy_levels_thin_new}))
for the inner radius $\rho_1=0.25$ (black lines) and  $\rho_1=5$ (red lines) as a function of continuous variable $m$. (Right) Three-dimensional representation of energy levels as a function of the quantum number $m$ considered as a continuous variable and the inner radius $\rho_1$.}
\label{fig:infinite-rho-2}
\end{figure}

\subsection{Solutions with zero angular momentum }

\subsubsection{Numerical analysis for a hole of an arbitrary size}

Now we analyze the energy levels distribution basing on Eq.~(\ref{energy_levels}) in the particular case of 
the quantum number $m=0$ and consider their dependence on 
the external radius $\rho_2$ with fixed internal radius $\rho_1$.  Figure ~\ref{fig:m=0-rho2} illustrates their characteristic deviations  from the standard Landau spectrum. One can see that for $\rho_1=0.25$ (left panel)
the energy levels tend to constant values that differ from the half integer 
Landau spectrum even for large $\rho_2$. This obviously is the consequence of the inner hole presence.
It will be shown below that the dispersion relation returns to the Landau spectrum with small corrections when $\rho_1 \to 0$, i.e. the central hole disappears.

For sufficiently large $\rho_1 >1$ the absence of the low energy levels (see the right panel of Fig.~\ref{fig:m=0-rho2})
reflects the fact that they sharply rise when $m=0$ (see Fig.~\ref{enery_level_general}d).
In other words, the flattening of the spectrum and approaching usual half-integer Landau levels takes place for sufficiently 
large $m$ only as clearly see from Fig.~\ref{enery_level_general}~(d). 
\begin{figure}
\includegraphics[width=0.49\columnwidth]{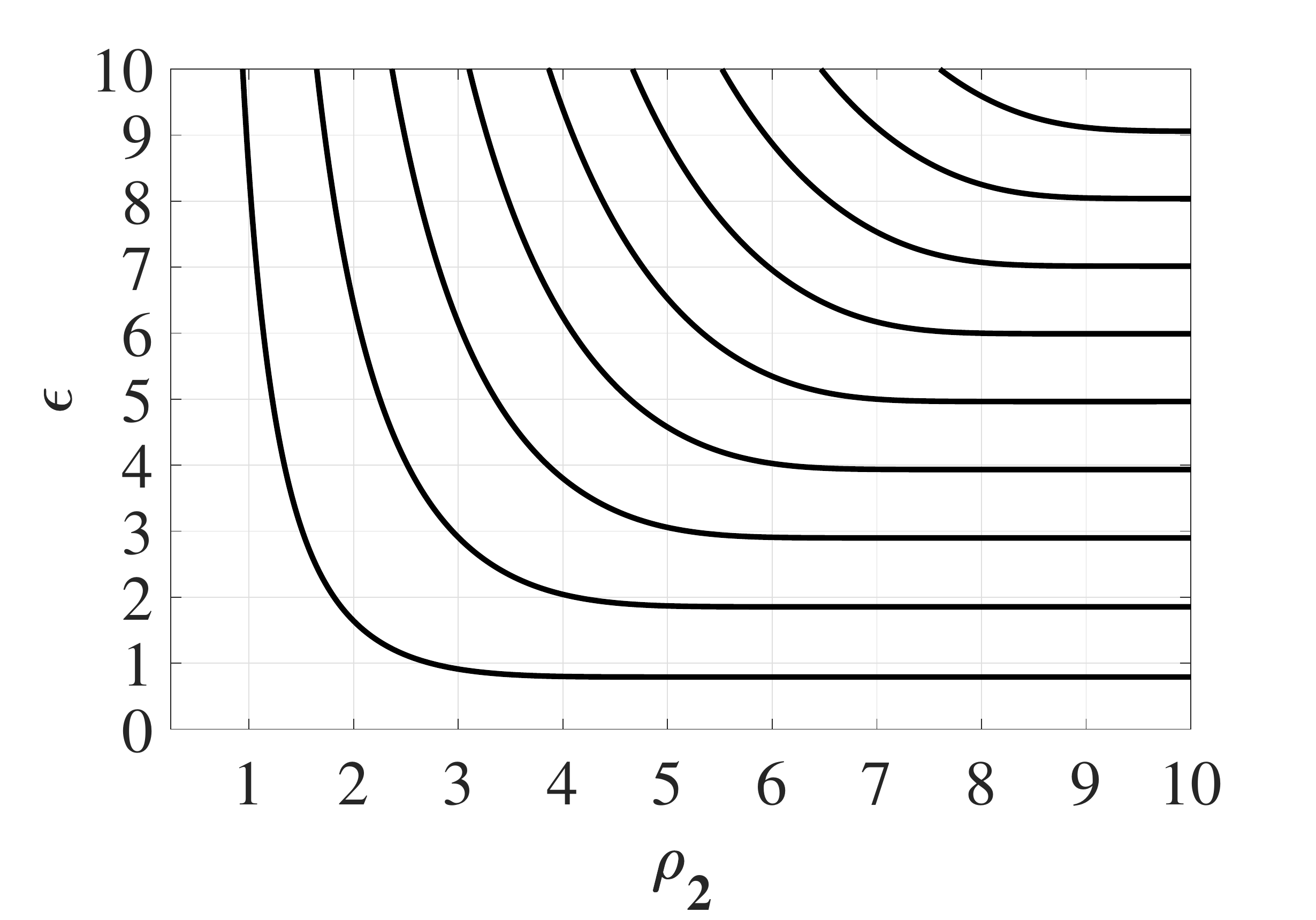}
\includegraphics[width=0.49\columnwidth]{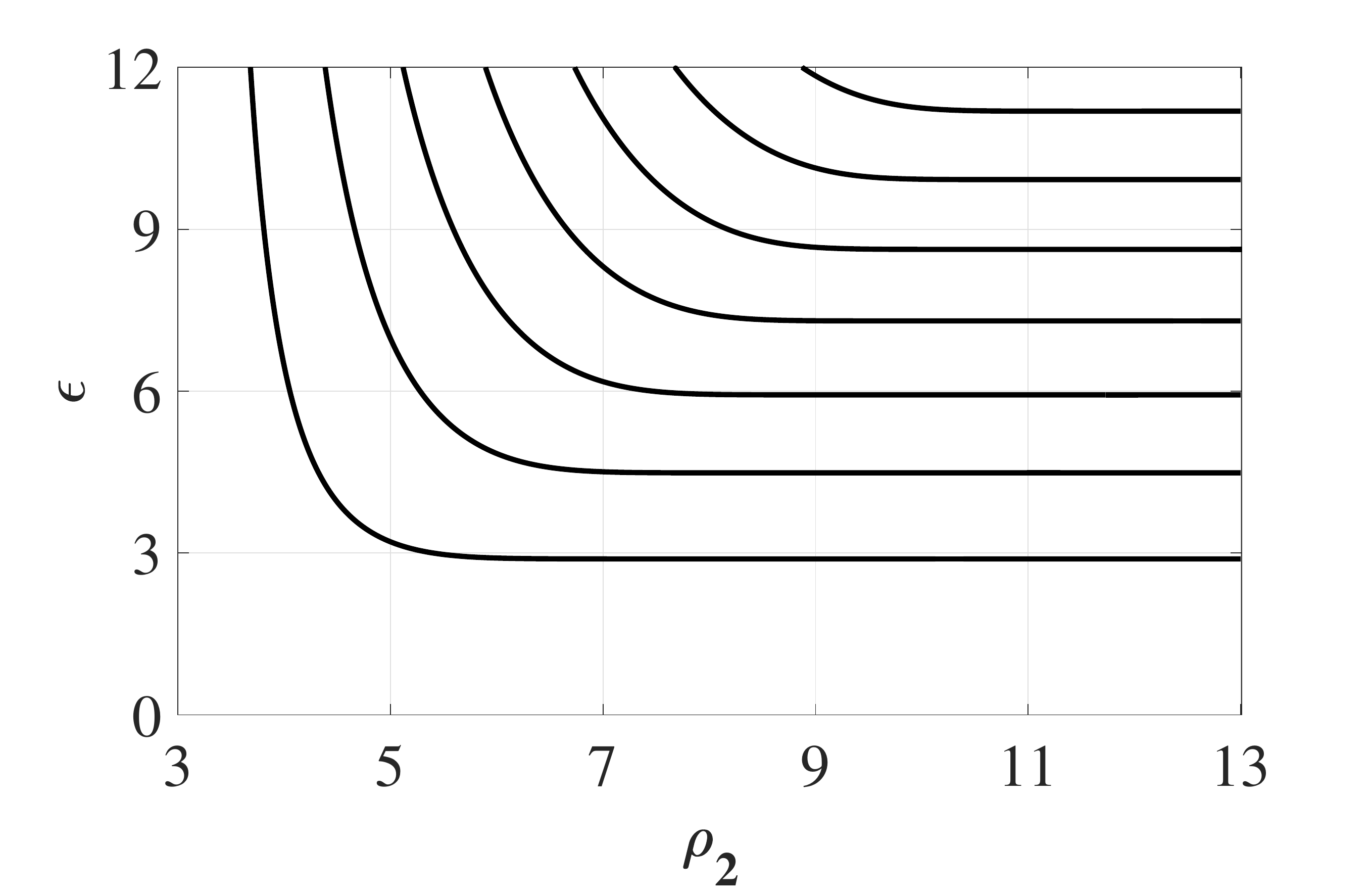}
\caption{The dependence of  of the energy levels with $m=0$ on the external radius $\rho_2$ 
for the two values of the internal radius:  $\rho_1 =0.25$ (left panel) and $\rho_1 =3$ (right).}
\label{fig:m=0-rho2}
\end{figure}

\subsubsection{The states with zero angular momentum in a disk with a small hole}

The mentioned above deviation of the spectrum from the standard Landau one can be found out analytically by considering  the specific limit $\rho_1 \ll 1$ and  $\rho_2 \to \infty$ (plane with a small hole). Using asymptotic expansions for the Whittaker function for the case  $m=0$ (i.e. $\tilde \varepsilon  = \varepsilon $)
with a small radius $\rho_1$ 
\begin{equation}
\label{assymptot_W_k=0}
{W_{\varepsilon ,0}}\left( {\frac{\rho ^2}{2}} \right) = \frac{1}{2}\frac{{\sqrt 2 \left( { - \psi \left( {\frac{1}{2} - \varepsilon } \right) - 2\gamma  + 2\ln \left( {\frac{1}{\rho }} \right) + \ln 2} \right)}}{{\Gamma \left( {\frac{1}{2} - \varepsilon } \right)}}\rho + O({\rho ^3}).
\end{equation}
one arrives at the transcendental equation with the digamma function $\psi(z)$:
\begin{equation}
\label{energy_levels_k=0}
- \psi \left( {\frac{1}{2} - \varepsilon } \right) - 2\gamma  + 2\ln \left( {\frac{1}{{{\rho _1}}}} \right) + \ln 2 = 0,
\end{equation}
which for $\rho_1 \ll 1$  gives the energy levels determined by the Eq. (\ref{energy_levels_m=0_new})
One can see that when $\rho_1 \to 0$ this spectrum tends to the standard Landau one, in complete agreement with the left panel of Fig.~\ref{fig:m=0-rho2}.

\subsection{Asymptotic of solutions with $m \to \infty $ }

Let us pass to the analysis of the energy levels with large angular momentum: $m\to\pm\infty$. We will do this basing on the same Eq.~(\ref{energy_levels_thin_new}), obtained in the assumption of $\rho_2 \gg 1$,  but do not requiring any more $\rho_1 \ll1$, i.e. we just fix $\rho_1 < \rho_2 $.

\subsubsection{General relations}

One can rewrite Eq.~(\ref{energy_levels_thin_new}) using the relation between the Whittaker function and the confluent hypergeometric function of the second kind $\Psi(a,b,z)$
\cite{Bateman1}:
\be
\label{hyper-rho1}
\Psi\left(\frac{1}{2}-\epsilon+\frac{|m|-m}{2},|m|+1; \frac{\rho_1^2}{2}\right)=0.
\ee
The last equation also follows directly from Eq.~(\ref{eq:confluent-functions}) in the limit $\rho_2 \to \infty$. It is valid for the states with arbitrary angular momentum, while for $m\ge0$ Eq.~(\ref{hyper-rho1}) reads as
\be
\Psi\left(\frac{1}{2}-\epsilon,m+1; 
m \lambda_m \right)=0.
\label{eq:through-r_k}
\ee
Here we introduced the parameter $\lambda_m= \rho_1^2/(2m)$.

The function $\Psi(a,b;x)$ for fixed $a$ and fixed $\lambda=x/b >0$  has the asymptotic expansion  given by Eq.~(13.8.5) from \cite{NIST}:
\be
\Psi(a,b;x)\sim b^{-\frac{a}{2}}e^{\frac{\zeta^2b}{4}}\left[\lambda\left(\frac{\lambda-1}{\zeta}\right)^{a-1}U(a-\frac{1}{2},\zeta\sqrt{b})
-\left(\lambda\left(\frac{\lambda-1}{\zeta}\right)^{a-1}-\left(\frac{\zeta}{\lambda-1}\right)^{a}\right)\frac{U(a-\frac{3}{2},\zeta\sqrt{b})}{\zeta\sqrt{b}}
\right]+O\left(\frac{1}{b}\right)
\label{expansionPsi-b,x-large}
\ee
for $b\to\infty$ uniformly in compact $\lambda$-intervals of $(0,\infty)$ and compact real $a$-intervals.
Here $\zeta=\sqrt{2(\lambda-1-\ln\lambda)}$ with ${\rm sign}(\zeta)={\rm sign}(\lambda-1)$, and the function $U(a, x)$ is related to the
parabolic cylinder function $U(a, x)=D_{-a-1/2}(x)$. The functions $U(a,x)$ and $U(a,-x)$ at large positive $x$ behave as
\ba
&&U(a,x)\simeq e^{-\frac{x^2}{4}}x^{-a-\frac{1}{2}}\left[1-\frac{(a+1/2)(a+3/2)}{2x^2}+O\left(\frac{1}{x^4}\right)\right],\quad x\to\infty,\nonumber\\
&&U(a,-x)\simeq\frac{\sqrt{2\pi}}{\Gamma(1/2+a)} e^{\frac{x^2}{4}}x^{a-\frac{1}{2}}-\sin(\pi a)
e^{-\frac{x^2}{4}}x^{-a-\frac{1}{2}},\quad x\to\infty,
\label{expansionU-large-x}
\ea
therefore we can neglect the second term in the expansion (\ref{expansionPsi-b,x-large}) and the equation (\ref{eq:through-r_k}) reduces to (note that
$\lambda<1$ and ${\rm sign}(\zeta)<0$)
\be
U(-\epsilon,-\zeta\sqrt{m})=0,\quad \zeta=\sqrt{2(\lambda-1-\ln\lambda)}>0.
\label{eq:large-k}
\ee

\subsubsection{The energy spectrum of skipping electrons}

Let us consider now the case when the center of wave function ($\rho_m$) is located  quite close to one of the edges:  $|2m-\rho_{1}|< 1$. 

i)  When  $|2m-\rho_{1}|=0$ one can rewrite Eq.~(\ref{hyper-rho1}), using the asymptotic expression for fixed $a$ and large $b$ in confluent hypergeometric function (see Eq.~(13.8.7) in \cite{NIST}):
\be
\Psi(a,b;b)=\sqrt{\pi}(2b)^{-\frac{a}{2}}\left[\frac{1}{\Gamma\left(\frac{a+1}{2}\right)}-\frac{(a+1)\sqrt{8/b}}{\Gamma\left(\frac{a}{2}\right)}
+O\left(\frac{1}{b}\right)\right],
\ee
that gives
\be
\frac{1}{\Gamma\left(\frac{3-2\epsilon}{4}\right)}-\frac{(3/2-\epsilon)\sqrt{8/m}}{\Gamma\left(\frac{1-2\epsilon}{4}\right)}=0.
\ee
Since $m\to\infty$, the energy levels are given by the poles of $\Gamma\left(\frac{3-2\epsilon}{4}\right)$.  which are at
$\epsilon=2n+3/2,\,n=0,1,2,\dots$. Solving the last equation we obtain the behavior of energy levels at edges of the disk:
\be
\varepsilon_{n,m}\simeq 2n+3/2+\frac{(3/2-2n)\Gamma(n+3/2)}{\pi n!}\sqrt{8/m},\quad m\to\infty.
\ee
 
ii) When $\rho_m$ is near the edge $\rho_{1}$ but $|2m-\rho_{1}|<1$ we consider Eq.(\ref{eq:large-k}) for $\lambda$ close to $1$
($\lambda\lesssim 1$) where $\zeta(\lambda)\simeq 1-\lambda$ and
\be
U(-\epsilon,(\lambda-1)\sqrt{m})=0.
\ee
Using the formulas (19.3.5) from [\onlinecite{Abramowitz}],
\be
U(a,0)=\frac{\sqrt{\pi}}{2^{\frac{a}{2}+\frac{1}{4}}\Gamma\left(\frac{3}{4}+\frac{a}{2}\right)},\quad
U'(a,0)=-\frac{\sqrt{\pi}}{2^{\frac{a}{2}-\frac{1}{4}}\Gamma\left(\frac{1}{4}+\frac{a}{2}\right)},\quad 
U''(a,0)=\frac{a\sqrt{\pi}}{2^{\frac{a}{2}+\frac{5}{4}}\Gamma\left(\frac{3}{4}+\frac{a}{2}\right)},
\ee
and keeping the terms up to $x^2$ in the expansion, we get the equation
\be
\frac{1+ax^2/2}{\Gamma\left(\frac{3-2\epsilon}{4}\right)}=\frac{\sqrt{2}x}{\Gamma\left(\frac{1-2\epsilon}{4}\right)},\quad x=(\lambda-1)\sqrt{m}.
\ee
As $x\to0$, the energies are given by poles of gamma function $\Gamma\left(\frac{3-2\epsilon}{4}\right)$ which are at $\epsilon=2n+3/2$, $n=0,1,\dots$.
Near the poles, writing $\epsilon=2n+3/2+\delta$ we have
\be
\frac{\Gamma\left(-n-1/2-\delta/2\right)}{\Gamma\left(-n-\delta/2\right)}= \sqrt{2}x +O\left(x^3\right).
\ee
Expanding in $\delta$ the equation reduces to
\be
\frac{\delta}{2}\left[1-\frac{\delta}{2}G(2n+2)\right]=\frac{\sqrt{2}\Gamma(n+3/2)}{\pi\Gamma(n+1)}x,
\ee
where $G(2z)=\psi(z+1/2)-\psi(z)$ is the known function and we used the relation for digamma function $\psi(-n-1/2)=\psi(n+3/2)$.
Thus we finally find
\be
\label{energies_hole_bulk_2}
\varepsilon_{n,m}= 2n+3/2+\frac{\Gamma(n+1/2)}{\sqrt{2}\pi n!}(2n+1)\frac{\rho_1^2-\rho_m^2}{\sqrt{m}}+G(2n+2)
\left(\frac{\Gamma(n+3/2)}{\pi n!}\right)^2\frac{(\rho_1^2-\rho_m^2)^2}{m}
\ee
[compare with Eq.~(7) in Ref.~\onlinecite{Heuser}]. We see the doubling of the frequency (with a shift) and the linear and quadratic corrections due to the edge.

\subsubsection{The energy spectrum of the electrons rotating on the cyclotron 
orbits far from the edge}

Now let us pass to the energy spectrum of electrons rotating on the cyclotron orbits 
far from the edges:
$|2m-\rho_{1}|>1$. Equation (\ref{expansionU-large-x}) gives
\ba
\frac{\sqrt{2\pi}}{\Gamma(1/2-\epsilon)}+\sin(\pi\epsilon)e^{-\frac{x^2}{4}}x^{2\epsilon}=0,\quad x=\zeta(\lambda)\sqrt{m},\quad \zeta(\lambda)=\sqrt{2(\lambda-1-\ln\lambda)}.
\ea
This is the same equation as for the energy spectrum of the states in the bulk of a ribbon \cite{Heuser,Varlamov} hence we can write
\be
\varepsilon_{n,m}=n+\frac{1}{2}+\frac{1}{\sqrt{2\pi}n!}x^{2n+1}e^{-x^2/2},\quad x=\zeta(\lambda)\sqrt{m},\,
\lambda=\frac{\rho_1^2}{\rho_m^2}<1.
\label{energies_hole_bulk}
\ee
and arrive at Eq.~(\ref{energy_levels_k=inf_bulk}).

\subsection{Back to Corbino disk}

Until now we analyzed the case of the infinite system with a hole. We checked that a similar consideration is valid
for the Corbino disk with the two edges described by the general equation for eigenvalues ~(\ref{eq:confluent-functions}). 
Using the asymptotic expression (\ref{expansionPsi-b,x-large}) for $\Psi$-functions and the analogous one for
\be
\Phi(a,b;\lambda b)\simeq b^{-\frac{a}{2}}e^{\frac{\zeta^2b}{4}}\left[\lambda\left(\frac{\lambda-1}{\zeta}\right)^{a-1}U(a-\frac{1}{2},-\zeta\sqrt{b})
+\left(\lambda\left(\frac{\lambda-1}{\zeta}\right)^{a-1}-\left(\frac{\zeta}{\lambda-1}\right)^{a}\right)\frac{U(a-\frac{3}{2},-\zeta\sqrt{b})}{\zeta\sqrt{b}}
\right]+O\left(\frac{1}{b}\right),
\label{expansionPhi-b,x-large}
\ee
for the fixed $\lambda_1=\rho_1^2/(2m)$ and $\lambda_2=\rho_2^2/(2m)$ Eq.~(\ref{eq:confluent-functions}) acquires
the form
\be
U(-\epsilon,\zeta_1\sqrt{m})U(-\epsilon,-\zeta_2\sqrt{m})-U(-\epsilon,-\zeta_1\sqrt{m})U(-\epsilon,\zeta_2\sqrt{m})=0,\quad \zeta_i=
\zeta(\lambda_i){\rm sign}(\lambda_i-1).
\ee

When $\rho_1<\sqrt{2m}<\rho_2$ we have $\lambda_1<1$ and $\lambda_2>1$, hence in this case we can write
\be
U(-\epsilon,-\zeta_1\sqrt{m})U(-\epsilon,-\zeta_2\sqrt{m})-U(-\epsilon,\zeta_1\sqrt{m})U(-\epsilon,\zeta_2\sqrt{m})=0,
\ee
with positive $\zeta_i=\sqrt{2(\lambda_i-1-\ln\lambda_i)}$. Thus, for $m\to\infty$, the equation splits in two:
\be
U(-\epsilon,-\zeta_1\sqrt{m})=0 \quad \mbox{and}\quad U(-\epsilon,-\zeta_2\sqrt{m})=0.
\ee
In the bulk, when $\rho_1<\sqrt{2m}<\rho_2$, where $\sqrt{2m}-\rho_1\gg 1$ and $\rho_2-\sqrt{2m}\gg 1$ (it is assumed that $\rho_1 \gg 1$ and $\rho_2-\rho_1 \gg 1$), the solutions of these equations give us the spectrum of electrons rotating along the cyclotron orbits (cp.  Eq.~(\ref{energies_hole_bulk})):
\be
\varepsilon_{n,m}= n+\frac{1}{2}+\frac{1}{\sqrt{2\pi}n!}x^{2n+1}e^{-x^2/2},\quad x=\zeta(\lambda_i)\sqrt{m},\,
\lambda_i=\frac{\rho_i^2}{\rho_m^2}.
\label{energies_disc_bulk_3}
\ee

The  energy spectra of the electrons skipping along the edges are determined by Eqs.~(\ref{energy_levels_k=inf_r1}) - (\ref{energy_levels_k=inf_r2}).

\section{Large and narrow Corbino disk}
\label{sec:B}

As one can see from Fig.~\ref{fig:m=0-rho2} (right panel) for a narrow Corbino disk with the large inner and outer radii energy tends to be large.  By means of the asymptotic expressions of Whittaker functions for large parameter $\epsilon$ (see Eqs.~(13.21.1) and (13.21.2) in \cite{NIST})
\begin{equation}
\label{M_large_parameter}
{M_{\tilde \varepsilon ,\frac{{\left| m \right|}}{2}}}\left( {\frac{1}{2}{\rho ^2}} \right) \approx \frac{\rho }{{\sqrt 2 }}\Gamma \left( {\left| m \right| + 1} \right){\tilde \varepsilon ^{ - \frac{{\left| m \right|}}{2}}}{J_{\left| m \right|}}\left( {\sqrt {2\tilde \varepsilon } \rho } \right),
\end{equation}
and
\begin{equation}
\label{W_large_parameter}
{W_{\tilde \varepsilon ,\frac{{\left| m \right|}}{2}}}\left( {\frac{1}{2}\rho _{}^2} \right) \approx \frac{\rho }{{\sqrt 2 }}\Gamma \left( {\tilde \varepsilon  + \frac{1}{2}} \right)
\left[ {\sin \left( {\pi \tilde \varepsilon  - \frac{{\pi \left| m \right|}}{2}} \right){J_{\left| m \right|}}\left( {\sqrt {2\tilde \varepsilon } \rho } \right) - \cos \left( {\pi \tilde \varepsilon  - \frac{{\pi \left| m \right|}}{2}} \right){Y_{\left| m \right|}}\left( {\sqrt {2\tilde \varepsilon } \rho } \right)} \right],
\end{equation}
we arrive at the new equation (\ref{energy_levels_Bessel})
for energy levels, which is much simpler than Eq.~(\ref{energy_levels}) or its equivalent Eq.~(\ref{eq:confluent-functions}). The asymptotics (\ref{M_large_parameter}) and 
(\ref{W_large_parameter}) are valid when the parameter $\rho^2/(8 \tilde \epsilon) <1$
(see the expansion (\ref{hypergeom_Bessel_expansion}) below and the relation between Whittaker and
confluent hypergeometric functions). 

Denoting in Eq.~(\ref{energy_levels_Bessel}) $x = \sqrt{2 \tilde \epsilon} \rho_1$
and $\lambda = \rho_2/\rho_1$ one can rewrite it in the form coinciding with
Eq.~(10.21.45) in \cite{NIST}
\begin{equation}
\label{Bessel-cross-equation}
J_{|m|}(x)Y_{|m|}(\lambda x)-J_{|m|}(\lambda x)Y_{|m|}(x)=0.
\end{equation}
Then for $\lambda >1$ and $\lambda -1 \ll 1$
one can take the first two terms of the asymptotic expansion for $n$th positive zeros
of the Bessel functions cross-product (see Eq.~(10.21.50) in \cite{NIST})
\begin{equation}
x_n =  \frac{\pi n}{\lambda-1} + 
\frac{4 m^2 -1}{8 \pi n}\frac{\lambda-1}{\lambda}, \qquad n=1,2, \ldots .  
\end{equation}
Accordingly, the solutions of Eq.~(\ref{Bessel-cross-equation}) are
\begin{equation}
\label{narrow-spectrum}
\tilde \epsilon = \frac{1}{2}\left[
\frac{\pi n}{\delta} + 
\frac{m^2-1/4}{2 \pi n}\frac{\delta}{\rho_1 \rho_2}
\right]^2,  \qquad n=1,2, \ldots ,     
\end{equation}
where $\delta = \rho_2 - \rho_1$. Restoring units 
and neglecting the term $\sim \delta^2$ ($d^2/r_1 r_2 \ll 1$)
one arrives at the final expression (\ref{narrow-spectrum-units}) for the spectrum of 
the narrow disk. The given above inequality $\rho^2/(8 \tilde \epsilon) <1$ results in the
condition $\delta \rho < 2 \pi$ which for $\rho > \delta$ implies that $\delta < \sqrt{2\pi}$.  

Using the large argument asymptotic of Bessel functions
(see Eqs.~(9.2.1) and (9.2.2) in \cite{Abramowitz}) one obtains from 
Eq.~(\ref{radial-wf_Bessel}) the following 
expression for the wave function
\begin{equation}
\label{asymptotic-wf}
f_{nm}(\rho)=\frac{\sin(\sqrt{2\tilde{\epsilon}}(\rho-\rho_1))}{l\sqrt{\pi\delta\rho}},
\end{equation}
where the normalization constant $C$ is determined by the condition (\ref{normalization-f}).

The formula (\ref{I_nm}) for the current $I_{nm}$ carried by the state with definite quantum numbers $n,m$ contains the following integral 
$\int_{\rho_1}^{\rho_2}d\rho f_{nm}^2(\rho)/\rho$. For large narrow Corbino 
disk we can use the found above asymptotic of the wave function (\ref{asymptotic-wf})
and obtain
\begin{equation}
\label{radial-integral}
\begin{split}
\int\limits_{\rho_1}^{\rho_2}\frac{d\rho}{\rho}f_{nm}^2(\rho)&=
\frac{\sqrt{2\tilde{\epsilon}}}{\pi l^2\delta}
\left[\frac{-1+\cos(2\sqrt{2\tilde{\epsilon}}\delta)}{2\sqrt{2\tilde{\epsilon}}\rho_2}+\sin(2\sqrt{2\tilde{\epsilon}}\rho_1)\left(
\mbox{Ci}(2\sqrt{2\tilde{\epsilon}}\rho_1)-\mbox{Ci}(2\sqrt{2\tilde{\epsilon}}\rho_2\right)\right.
\\
&-\left.\cos(2\sqrt{2\tilde{\epsilon}}\rho_1)\left(
\mbox{Si}(2\sqrt{2\tilde{\epsilon}}\rho_1)-\mbox{Si}(2\sqrt{2\tilde{\epsilon}}\rho_2)\right)\right],
\end{split}
\end{equation}
where $\mbox{Si}(z)$ and $\mbox{Ci}(z)$ are sine integral and cosine integral functions, respectively. Using their asymptotic at large argument,
\begin{equation}
\mbox{Si}(z)\approx \frac{\pi}{2}-\frac{\cos z}{z},
\qquad \mbox{Ci}(z)\approx \frac{\sin z}{z},
\qquad z \gg 1,
\end{equation}
and that in the leading approximation from Eq.~(\ref{narrow-spectrum}) follows that $\sqrt{2\tilde{\epsilon}}\delta=\pi n$, we arrive at the following simple result
\begin{equation}
\label{integral}
\int\limits_{\rho_1}^{\rho_2}\frac{d\rho}{\rho}f_{nm}^2(r)=\frac{1}{2\pi l^2\rho_1\rho_2}.
\end{equation}

\section{Small Corbino disk with infinitesimal inner radius}
\label{sec:C}

In this case one can approximate a Corbino disk as a solid disk without a hole in the centre with the conditions ${\rho _2} < 1$  and ${\rho _1}=0$. The solution for energy spectrum and is given by zeroes of the confluent hypergeometric function (\ref{energy_levels_small_disc}) (see e.g. \cite{Rensink}).
We rewrite this expression by means of the formula that represents the confluent hypergeometric function as the series of the Bessel functions of the first kind (see Eq.~(13.3.7) in \cite{Abramowitz})
\begin{equation}
\label{hypergeom_Bessel_expansion}
\Phi \left( {\frac{1}{2} - \varepsilon  + \frac{{\left| m \right| - m}}{2},\left| m \right| + 1,\frac{1}{2}\rho _2^2} \right) =
\Gamma(|m|+1)
{e^{\frac{1}{4}\rho _2^2}}{\tilde \varepsilon ^{ - \frac{1}{2}\left| m \right|}}\mathop \sum \limits_{p = 0}^\infty  {A_p}{
\left( \frac{\rho _2^2}{ 8 \tilde \varepsilon} \right)^{\frac{p}{2}}}
{J_{\left| m \right| + p}}\left( {\sqrt {2\tilde \varepsilon } {\rho _2}} \right),
\end{equation}
where coefficients satisfy the recurrence relation 
\begin{equation}
\label{coefficients_expansion}
\left( {n + 1} \right){A_{n + 1}} = \left( {n + \left| m \right|} \right){A_{n - 1}} - 2\tilde \varepsilon {A_{n - 2}},
\end{equation}
and ${A_0} = 1$, ${A_1} = 0$, ${A_2} = \frac{1}{2}\left| m \right| + \frac{1}{2}$.

If the expansion parameter $\rho_2/(2 \sqrt{2 \tilde \varepsilon} )$ is small then we can keep the first term in 
Eq.~(\ref{hypergeom_Bessel_expansion}).  
This leads to the  equation for eigenvalues
\begin{equation}
\label{spectrum_small_disc_Bessel}
{J_{\left| m \right|}}\left( {\sqrt {2\tilde \varepsilon } {\rho _2}} \right) = 0,
\end{equation}
which has the solutions $\sqrt {2\tilde \varepsilon } \rho_2 = j_{nm}$
with ${j_{nm}}$ being the n-th root of
the equation ${J_{\left| m \right|}}\left( z \right) =0$.
The corresponding energy spectrum $E_{n,m}$ is 
given by 
Eq.~(\ref{energy_levels_small_disc_new}) in the main text.
The expansion parameter  $\rho_2/(2 \sqrt{2 \tilde \varepsilon}) <1$
in Eq.~(\ref{hypergeom_Bessel_expansion})
becomes
$\rho_2^2/(2 j_{nm}) < \rho_2^2/(2 j_{10})$.
Thus, using the value of the lowest root  $j_{10} \approx 2.4$, we estimate that $r_2 < 2.2 l$
which determines the range of validity of the considered approximation.
Note that Eq.~(\ref{spectrum_small_disc_Bessel}) also follows directly
from Eq.~(\ref{energy_levels_Bessel}) by taking there $\rho_1 =0$.

In turn, the radial component of the wave function has the form
\begin{equation}
f_{nm}\left( r \right) = C{J_{\left| m \right|}}\left( {\sqrt {2\tilde \varepsilon } \rho } \right) = C J_{|m|} 
\left( j_{nm} \frac{r}{r_2} \right).
\end{equation}
The constant $C$ can be found from the normalization condition (\ref{normalization-f})
(see Eq.~(6.521.1) in \cite{Gradstein}):
\begin{equation}
\label{C_small_Corbino_disc}
{C^2} = \frac{1}{\pi {r_2^2J_{\left| m \right| + 1}^2\left( {{j_{nm}}} \right)}}.
\end{equation}
Therefore, the radial component of the wave function is given by 
Eq.~(\ref{radial_wave_func_small_disc}) in the main text.
In this case the full current given by 
Eqs.~({\ref{full_current}}) and (\ref{I_nm})
acquires the following form
\begin{equation}
\label{prel_current_small_disc}
I =  \frac{{ e\hbar }} {2 \pi {{m_e} r_2^2}}\sum_{\substack{m = -\infty\\ n=1 } }^\infty 
\frac{\theta(\mu- E_{n,m}) }{J_{\left| m \right| + 1}^2\left( {{j_{nm}}} \right)}
{\int\limits_0^{{r_2}} dr { \left( {  \frac{{2m}}{r } - \frac{r}{l^2} } \right)} } 
J_{\left| m \right|}^2\left( {{j_{nm}}\frac{r }{{{r_2}}}} \right) .  
 \end{equation}
The integration of the first term in the bracket can be done using the following formula (Eq.~(1.8.3.17) in \cite{Prudnikov2})
\begin{equation}
\label{A_nm}
A_{nm} \equiv
2|m| \int\limits_{0}^{1}\frac{dr}{r}J^2_{|m|}(j_{nm} r)=1+J_0^2(j_{nm})-2\sum\limits_{k=0}^{|m|-1}J^2_k(j_{nm}),\qquad |m|\ge1,
\end{equation}
while the second term is integrated using the
normalization condition (\ref{normalization-f}). Thus we arrive at 
Eq.~(\ref{current_small_disc_main}) in the main text.

To verify that Eqs.~(\ref{current_small_disc_main}) -
(\ref{current_small_disc_main_2})  follow directly from
the Byers-Yang formula (\ref{Byers-Yang}) one needs to calculate explicitly
the derivative  $\partial j_{n \nu}/ \partial \nu $ of the roots $j_{n\nu}$ of the equation
$J_{\nu}(j_{n\nu})=0$ with $\nu = |m|$.
It can be expressed as follows [see Ref.~\onlinecite{Watson-Bessels-book} Section 15.6, Eq.~(2)]:
\begin{equation}
\frac{\partial j_{n\nu}}{\partial\nu}=\frac{2\nu}{j_{n\nu}J^2_{\nu+1}(j_{n\nu})}\int\limits_0^1\frac{dx}{x}J^2_\nu(j_{n\nu}x).
\end{equation}
Accordingly, one obtains 
\begin{equation}
\label{root-bessel-derivative}
\frac{\partial j_{nm}}{\partial|m|}=
\frac{2|m|}{j_{nm}J^2_{|m|+1}(j_{nm})}
\int\limits_0^1\frac{dz}{z}J^2_{|m|}(j_{n\nu}z)=\frac{A_{nm}}{j_{nm}J^2_{|m|+1}(j_{nm})},
\end{equation}
where in the last identity the definition
(\ref{A_nm}) for $A_{nm}$ is taken into account. 

\end{widetext}

\end{document}